\documentclass[preprintnumbers,article,amsmath,amssymb,floatfix,10pt,prd,onecolumn,superscriptaddress,nofootinbib]{revtex4-2}
\usepackage{titlesec}

\titleformat*{\section}{\LARGE\bfseries}
\titleformat*{\subsection}{\Large\bfseries}
\titleformat*{\subsubsection}{\large\bfseries}
\titleformat*{\paragraph}{\large\bfseries}
\titleformat*{\subparagraph}{\large\bfseries}

\usepackage{bm}
\usepackage{amsfonts}
\usepackage{latexsym}
\usepackage[latin1]{inputenc}
\usepackage{graphicx}
\usepackage{amsmath}
\usepackage{palatino}
\usepackage{mathpazo}
\usepackage{textcomp}
\usepackage{float}
\usepackage{booktabs}
\usepackage{dcolumn}
\usepackage{ragged2e}
\usepackage{hyperref}
\hypersetup{colorlinks,citecolor=blue}
\hypersetup{colorlinks=true,linkcolor=red,filecolor=magenta,urlcolor=blue}
\usepackage{amsmath}
\usepackage{xcolor}
\usepackage{orcidlink}
\usepackage{epsfig}
\usepackage[caption=false]{subfig}
\usepackage{commath}
\usepackage{cancel, ulem}
\usepackage{multirow}

\newcommand{\m}{\mathring}
\newcommand{\be}{\begin{equation}}
\newcommand{\ee}{\end{equation}}
\newcommand{\bea}{\begin{eqnarray}}
\newcommand{\eea}{\end{eqnarray}}
\newcommand{\eeas}{\end{eqnarray*}}
\newcommand{\beas}{\begin{eqnarray*}}

\def\jnl@style{\it}
\def\aaref@jnl#1{{\jnl@style#1}}

\def\aaref@jnl#1{{\jnl@style#1}}

\def\aj{\aaref@jnl{AJ}}                   
\def\apj{\aaref@jnl{ApJ}}                 
\def\apjl{\aaref@jnl{ApJ}}                
\def\apjs{\aaref@jnl{ApJS}}               
\def\apss{\aaref@jnl{Ap\&SS}}             
\def\aap{\aaref@jnl{A\&A}}                
\def\aapr{\aaref@jnl{A\&A~Rev.}}          
\def\aaps{\aaref@jnl{A\&AS}}              
\def\mnras{\aaref@jnl{Mon.~Not.~Roy.~Astron.~Soc.}}             
\def\prd{\aaref@jnl{Phys.~Rev.~D}}        
\def\prc{\aaref@jnl{Phys.~Rev.~C}}  
\def\prl{\aaref@jnl{Phys.~Rev.~Lett.}}    
\def\qjras{\aaref@jnl{QJRAS}}             
\def\skytel{\aaref@jnl{S\&T}}             
\def\ssr{\aaref@jnl{Space~Sci.~Rev.}}     
\def\zap{\aaref@jnl{ZAp}}                 
\def\nat{\aaref@jnl{Nature}}              
\def\aplett{\aaref@jnl{Astrophys.~Lett.}} 
\def\apspr{\aaref@jnl{Astrophys.~Space~Phys.~Res.}} 
\def\physrep{\aaref@jnl{Phys.~Rep.}}      
\def\physscr{\aaref@jnl{Phys.~Scr}}       
\def\commat{\aaref@jnl{Comm.~Math.~Phys.}}              
\def\science{\aaref@jnl{Science}}               
\def\cqg{\aaref@jnl{Classical Quant.~Grav.}}            
\def\jpcs{\aaref@jnl{JPCS}}                                     
\def\ijmpd{\aaref@jnl{Int.~J.~Mod.~Phys.~D}}                    
\def\grg{\aaref@jnl{Gen.~Relat.~Gravit.}}               
\def\rpp{\aaref@jnl{Rep.~Prog.~Phys.}}          
\def\npa{\aaref@jnl{Nucl.~Phys.~A}}        
\def\lrr{\aaref@jnl{Living Rev.~Rel.}}                   
\def\jcap{\aaref@jnl{J.~Cosmology Astropart.~Phys.}}    
\def\rmp{\aaref@jnl{Rev.~Mod.~Phys.}}   
\def\epjc{\aaref@jnl{Eur.~Phys.~J.~C}} 
\def\plb{\aaref@jnl{~Phy.~Lett.~B}} 
\def\mpla{\aaref@jnl{Mod.~Phy.~Lett.~A}} 
\def\arxiv{\aaref@jnl{arxiv.org}}

\allowdisplaybreaks[1]

\begin{document}

\title{How different connections in flat FLRW geometry impact energy conditions in $f(Q)$ theory? }

\author{Ganesh Subramaniam\orcidlink{0000-0001-5721-661X}}
\email{ganesh03@1utar.my}
\address{Department of Mathematical and Actuarial Sciences\\
Universiti Tunku Abdul Rahman, Jalan Sungai Long, 43000 Cheras, Malaysia}

\author{Avik De\orcidlink{0000-0001-6475-3085}}
\email{avikde@utar.edu.my}
\address{Department of Mathematical and Actuarial Sciences\\
Universiti Tunku Abdul Rahman, Jalan Sungai Long, 43000 Cheras, Malaysia}

\author{Tee-How Loo\orcidlink{0000-0003-4099-9843}}
\email{looth@um.edu.my}
\address{Institute of Mathematical Sciences, Faculty of Science\\
Universiti Malaya, 50603 Kuala Lumpur, Malaysia}

\author{Yong Kheng Goh\orcidlink{0000-0002-7338-9614}}\email{gohyk@utar.edu.my}
\address{Department of Mathematical and Actuarial Sciences\\
Universiti Tunku Abdul Rahman, Jalan Sungai Long, 43000 Cheras, Malaysia}

\footnotetext{The research was supported by the Ministry of Higher Education (MoHE), through the Fundamental Research Grant Scheme (FRGS/1/2021/STG06/UTAR/02/1). }

\begin{abstract}
In this study of the modified $f(Q)$ theory of gravity in the spatially flat Friedmann-Lema\^itre-Robertson-Walker (FLRW) spacetime, we explore all the affine connections compatible with the symmetric teleparallel structure; three classes of such connections exist, each involving an unknown time-varying parameter $\gamma(t)$. Assuming ordinary barotropic fluid as the matter source, we first derive the Friedmann-like pressure and energy equations. Next we impose constraints on the parameters of two separate $f(Q)$ models, $f(Q)=Q+\beta Q^2$ and $f(Q)=Q+\beta\sqrt{-Q}$ from the traditional energy conditions. Observational values of some prominent cosmological parameters are used for this purpose, yielding an effective equation of state $\omega^{eff}=-0.82$.   
\end{abstract}

\maketitle
\tableofcontents
\section{Introduction}\label{sec1}
Energy conditions (EC), first introduced in the classical general relativity (GR), 
were initially popularised by Penrose and Hawking when they applied these conditions to understand the singularity formed by gravitational collapse by imposing non-negativity to local energy, singularity during the beginning of the Universe and the causal structure of the Universe \cite{penrose1965, hawking1966, hawking1966i, hawking1966ii}. It has been understood that singularity is unavoidable in a spacetime satisfying null energy condition (NEC) ($\rho+p\geq0$) and strong energy condition (SEC) ($\rho+3p\geq0$) \cite{hawking1969, hawking1970} (see \cite{hawking201} for a more comprehensive treatment). ECs, in addition to playing an important role in understanding big bang singularity and black hole singularity, are used to prove several theorems, such as topological censorship \cite{friedman1993}, proof of the second law of black hole thermodynamics \cite{visser1996}, and the Hawking area theorem \cite{hawking1971}, which was also observationally verified \cite{isi2021}.
Apart from that, the weak energy condition (WEC) could be violated in an eternally inflating spacetime, and it is unclear whether a satisfactory model free of initial singularity could be constructed \cite{borde1997}. In contrast, the authors of \cite{kontou2021} demonstrated how eternal inflation is possible without violating the ECs.

For a perfect fluid type stress-energy tensor $T^m_{\mu\nu}=(p^m+\rho^m)u_\mu u_\nu+p^mg_{\mu\nu}$, we can derive the very simple and popularmost form of energy conditions
\begin{itemize}
    \item Weak energy condition (WEC): $\rho^{m}\ge0$, $\rho^{m}+p^{m}\ge0$.
    \item Null energy condition (NEC): $\rho^{m}+p^{m}\ge0$.
    \item Dominant energy condition (DEC): $\rho^{m}\pm p^{m}\ge0$.
    \item Strong energy condition (SEC): $\rho^{m}+3p^{m}\ge 0$, $\rho^{m}+p^{m}\ge0$. 
\end{itemize}

If we restrict ourself in the most supported (both theoretically and observationally) model of the current homogeneous and isotropic universe, that is, the spatially flat Friedmann-Lema\^itre-Robertson-Walker (FLRW) spacetime
\begin{equation}\label{metric}
ds^2=-dt^2+a^2(t)\left(dr^2+r^2d\theta^2+r^2\sin^2\theta d\phi^2\right),
\end{equation}
and give a closer look at the corresponding Friedmann equations of GR
(we, without loss of generality, consider the Einstein gravitational constant $\kappa=1$ for the time being)
\begin{subequations}\label{1a}
\begin{align}
    \rho^m&=3H^2\,,\\
    p^m&=-(2\dot{H}+3H^2),
\end{align}
\end{subequations}
we find that for the current accelerating state of the universe, that is ($a>0,\, \dot{a}>0$ as well as $\ddot{a}>0$), all the energy conditions except SEC fit in beautifully.
Here and in the remaining text, the $\dot{(~)}$ denotes the derivative over $t$.
The condition
$\rho^m$ is obviously non-negative as it equals a perfect square term, $\rho^m+p^m=-2\dot{H}=+2(q+1)H^2\ge0$ and $\rho^m-p^m=2\dot{H}+6H^2=2\frac{\ddot{a}}{a}+4H^2\ge0$,
where $H$ and $q$ denote  the Hubble parameter and deceleration parameter respectively 
(cf. (\ref{eqn:H})--(\ref{eqn:q})). 
On the other hand, $\rho^m+3p^m=-6(\dot{H}+H^2)=-6\frac{\ddot{a}}{a}<0$. How to justify this? We attribute this to the presence of an exotic matter with negative pressure ($p^{DE}<0$, to be precise $p^{DE}<-2\frac{\ddot{a}}a$) reasoning a violation of the SEC in case of the accelerating expansion. Let us modify (\ref{1a}) and rewrite as
\begin{subequations}\label{2a}
\begin{align}
    \rho^m&=3H^2\,,\\
    p^m+p^{DE}&=-(2\dot{H}+3H^2).
\end{align}
\end{subequations}
WEC, NEC, SEC now fit in perfectly. Little trouble arises with DEC, that is, non-negativity of 
\begin{align}
\rho^m-p^m=2(\dot{H}+3H^2)+p^{DE}\,,
\end{align}
recalling that $p^{DE}$ is actually negative, however, with a suitable lower bound ($p^{DE}>-2(2-q)H^2$), we can abolish the issue. Combining, we come to a conclusion that the physical presence of an exotic matter or dark energy in the universe whose pressure satisfies $-2(2-q)H^2<p^{DE}<-2\frac{\ddot{a}}a$ can explain the present accelerating expansion of the universe in the realm of GR.

Unfortunately, such an exotic matter is yet undetected which motivated academics to look for other alternate explanations. Modification of GR is one such alternative, which allows us to describe the present accelerated expansion of the Universe without needing a dark energy. In modified gravity theories, we change the Einstein-Hilbert action of GR suitably using the geometry of the underlying spacetime which in turn produces an additional geometric component. These additional terms generated from the modification of the theory itself can act as a negative pressure component and drive the acceleration. Ordinary matter always satisfies all ECs, and this fact is gravity theory independent. So even if we modify the geometric side of the field equation in GR and consider any modified theory models like $f(R)$, $f(Q)$, $f(T)$ etc; the right hand side, that is, the ordinary matter is still bound by the ECs. Thus we can utilise the ECs to obtain permitted ranges of the model parameters of such modified $f(.)$ theories of gravity.  

The beyond-GR studies of EC always remain a bit delicate. EC constraints in several modified gravity theories have been carried out, for example, in $f(R)$ theory \cite{bergliaffa2006, santos2007, bertolami2009}, $f(G)$ theory \cite{13, bamba2017}, $f(T)$ theory \cite{lie2012}, $f(G,T)$ theory \cite{sharif2016}, $f(R,T,R_{\mu\nu}T_{\mu\nu})$ theory \cite{sharif2013}, $f(R,G)$ theory \cite{18}, $f(R,\Box R,T)$ gravity \cite{19}, $f(R,T)$ theory \cite{20}, among others.
In fact, a study of ECs in any newly proposed gravity theory acts as an effective filtration to choose correct model parameters. 

Recently, teleparallel gravity (TG) as Einstein's other theory of gravity, is studied widely where the ``metric-compatible and torsion-free'' Levi Civita connection is replaced by a more general affine connection in a flat spacetime. There are two typical types of teleparallelism available in the literature, the first is the torsion-based metric telepallelism and the second is the symmetric teleparallelism in a torsion-less environment where the non-vanishing non-metricity tensor drives the gravity. In the present study, we concentrate on the $f(Q)$ theories of gravity, newly-proposed in the symmetric teleparallelism to avoid the DE-dependencies of GR.

In the $f(Q)$ theory, the ECs have been studied in \cite{mandal2020}, however, the authors have used the special FLRW line element in Cartesian coordinates, the coincident gauge in this setting. It has made their calculation easier by reducing the covariant derivative to merely a partial derivative, however, the Friedmann equations are identical to those of the $f(T)$ theories and therefore the purpose of investigating a new gravity theory remained unfulfilled. Moreover, there is reasonable hardcore mathematical evidence showing that their effective pressure and energy density expressions have been missing some crucial terms \cite{de/comment}. Hence, a complete study of ECs for all the possible affine connections compatible with the $f(Q)$ theory in the FLRW spacetime, is very important for future advancement of this newly proposed gravity theory in the correct path. The previously studied case can also be scrutinised from our analysis as a simple corollary.  
 
The present article is structured as follows:
After the introductory texts, in Section \ref{sec2} we offer a comprehensive mathematical formulation of the symmetric teleparallelism, specially the $f(Q)$ theory, and all the possible affine connections in the spatially flat FLRW spacetime which can yield flatness and torsionless geometry yet a non-zero non-metricity tensor. There are three classes of such connections available, discussed in three separate subsections \ref{sec2-1}, \ref{sec2-2}, \ref{sec2-3}. Each of these three connection classes involves an unknown and so far unconstrained time-varying parameter $\gamma(t)$. We recall two of the most prominent temporal functions of modern cosmology, the scale factor $a(t)$ and the Hubble parameter $H(t)=\frac{\dot{a}}{a}$ containing its first time derivative. To investigate the role of $\gamma(t)$ further, we have considered some reasonable ansatz, $\gamma(t)=\gamma_0$, $\gamma(t)=\gamma_1 a(t)$ and $\gamma(t)=\gamma_2 H(t)$, $\gamma_0, \, \gamma_1, \, \gamma_2$ being constants, to be examined from the perspective of ECs next. Additionally, we have studied a special case in each of Connection class II and III respectively, $\gamma(t)=\frac{\gamma_3}{a^{3}}$ and $\gamma(t)=\frac{\gamma_4}{a}$.

Since $\gamma(t)$ is non-vanishing in connection class II and III, the proportionality constants are non-zero. The respective pressure and energy equations for ordinary matter as well as the effective fluid are presented therein. Two popular $f(Q)$ models, namely, $f(Q)=Q+\beta Q^2$ and $f(Q)=Q+\beta \sqrt{-Q}$ are considered and the range of ($\beta,\gamma$) for valid ECs are computed for each class of connections, accompanied by contour plots for easy access. Finally, in Section \ref{sec3} we conclude our findings. 

For the quadratic model, the unit of $\beta$ has to be $\frac{Mpc^2}{km^2/s^2}$ which is the unit of $\frac{1}{H^2}$ and for squareroot model the $\beta$ unit has to be $km/s/Mpc$ which is the unit of $H$. Furthermore, $\gamma, \gamma_0$, $\gamma_1$, $\gamma_3$ and $\gamma_4$ has the unit of $km/s/Mpc$ which is the unit of $H$ while $\gamma_2$ is a dimensionless quantity.

The notations  $f_Q=\frac{df}{dQ}$ and $f_{QQ}=\frac{d^2f}{dQ^2}$ are used. Since $\gamma(t)$ is non-vanishing in connection class II and III, and further to reduce complexity of the analysis, we consider only positive range of $\gamma$.
\newpage

\section{Symmetric teleparallelism in spatially flat FLRW universe}\label{sec2}
Corresponding to the metric structure $g_{\mu\nu}$ in a spacetime, there is a special unique connection which is torsion-free and is also compatible with it. This connection goes under the name of Levi-Civita connection and is given by
\begin{equation}
\mathring{\Gamma}^\alpha_{\,\,\,\mu\nu}=\frac{1}{2}g^{\alpha\beta}\left(\partial_\nu g_{\beta\mu}+\partial_\mu g_{\beta\nu}-\partial_\beta g_{\mu\nu}  \right).
\end{equation}

In general, symmetric teleparallel gravity theory is constructed from a general affine connection $\Gamma^\alpha_{\,\,\, \beta\gamma}$ 
having 
null curvature and torsion, and its non-vanishing non-metricity tensor alone controls the gravity. We define the non-metricity tensor 
\begin{equation} \label{Q tensor}
Q_{\lambda\mu\nu} = \nabla_\lambda g_{\mu\nu}.
\end{equation}
The two independent traces of this non-metricity tensor are 
\[
Q_{\lambda}=Q_{\lambda\mu\nu}g^{\mu\nu}; \quad \tilde Q_{\nu}=Q_{\lambda\mu\nu}g^{\lambda\mu},
\]
which enter the definition of the disformation tensor $L^\lambda{}_{\mu\nu}$ and the non-metricity conjugate tensor $P^\lambda{}_{\mu\nu}$ 
\begin{equation} \label{L}
L^\lambda{}_{\mu\nu} = \frac{1}{2} (Q^\lambda{}_{\mu\nu} - Q_\mu{}^\lambda{}_\nu - Q_\nu{}^\lambda{}_\mu) \,,
\end{equation}
\begin{equation} \label{P}
P^\lambda{}_{\mu\nu} = \frac{1}{4} \left( -2 L^\lambda{}_{\mu\nu} + Q^\lambda g_{\mu\nu} - \tilde{Q}^\lambda g_{\mu\nu} -\frac{1}{2} \delta^\lambda_\mu Q_{\nu} - \frac{1}{2} \delta^\lambda_\nu Q_{\mu} \right) \,.
\end{equation}
Then the affine connection can be decomposed as 
\begin{equation} \label{connc}
\Gamma^\lambda{}_{\mu\nu} = \mathring{\Gamma}^\lambda{}_{\mu\nu}+L^\lambda{}_{\mu\nu}\,.
\end{equation}
Finally, we construct the non-metricity scalar  
\begin{equation} 
Q=Q_{\lambda\mu\nu}P^{\lambda\mu\nu}= \frac{1}{4}(-Q_{\lambda\mu\nu}Q^{\lambda\mu\nu} + 2Q_{\lambda\mu\nu}Q^{\mu\lambda\nu} +Q_\lambda Q^\lambda -2Q_\lambda \tilde{Q}^\lambda).
\end{equation}
Note that, in literature there exists another definition $Q=-Q_{\lambda\mu\nu}P^{\lambda\mu\nu}$, which gives a change of sign in the non-metricity scalar $Q$. This is important to keep in mind while comparing different results in $f(Q)$ articles.

Now, one can replace the Ricci scalar $R$ by this non-metricity scalar $Q$ in the Einstein-Hilbert action of GR to produce symmetric teleparallel equivalent of GR (STEGR). However, being equivalent to GR, the symmetric teleparallel theory inherits the same `dark' problem as in GR, and so a modified $f(Q)$-gravity has been introduced in the same way as a modified $f(R)$-theory was introduced to extend GR. By varying the action term 
\begin{equation*}
S = \frac1{2\kappa}\int f(Q) \sqrt{-g}\,d^4 x
+\int \mathcal{L}_M \sqrt{-g}\,d^4 x\,,
\end{equation*}
with respect to the metric, we obtain the field equation
\begin{equation} \label{FE1}
\frac{2}{\sqrt{-g}} \nabla_\lambda (\sqrt{-g}f_QP^\lambda{}_{\mu\nu}) -\frac{1}{2}f g_{\mu\nu} + f_Q(P_{\nu\rho\sigma} Q_\mu{}^{\rho\sigma} -2P_{\rho\sigma\mu}Q^{\rho\sigma}{}_\nu) = \kappa T^{\text m}_{\mu\nu}.
\end{equation}
Recently, the covariant formulation of this field equation was obtained and used effectively to study the geodesic deviations and in cosmological sector \cite{zhao,gde}
\begin{equation} \label{FE}
f_Q \m{G}_{\mu\nu}+\frac{1}{2} g_{\mu\nu} (f_QQ-f) + 2f_{QQ} P^\lambda{}_{\mu\nu} \m{\nabla}_\lambda Q = \kappa T^{\text m}_{\mu\nu}.
\end{equation}
Several important publications came up very recently on this modified $f(Q)$-gravity theory and its cosmological implications, see \cite{coincident,cosmology,cosmography,lu,lin, de/isotropization,de/accelerating,de/complete,cosmology_Q,redshift,signature,lcdm1,siren,ambrosio2022,anagnostopoulos2021,lazkos2019,arora2022,gadbail2022,solanki2021,harko2018,bajardi2020,lymperis2022,dixit2022,kyllep2022,chanda2022,sahoo2022,aziza2021,de2022,dambrosio2022} and the references therein. However, all these studies were conducted in the coincident gauge choice, line element in Cartesian coordinates. This specific choice reduced the covariant derivative into partial derivative, making the calculations simpler. 
But at the adverse side, the energy and pressure equations are identical with the $f(T)$ theory. 
Additionally, this particular affine connection is merely the special case of one of the three classes to be discussed in the next paragraph (more precisely, the connection of type I  with $\gamma=0$). 
So we must look beyond this particular gauge choice.   

In the current discussion, we consider all the compatible connections, namely, the torsion free connections with zero curvature, in the spatially flat FLRW spacetime metric (\ref{metric}). There are total three possible classes of such connections, depending on the purely temporal functions $C_1$, $C_2$ and $C_3$ given by \cite{FLRW/connection,FLRW/connection1}
\begin{align} 
\Gamma^t{}_{tt}=&C_1, 
	\quad 					\Gamma^t{}_{rr}=C_2, 
	\quad 					\Gamma^t{}_{\theta\theta}=C_2r^2, 
	\quad						\Gamma^t{}_{\phi\phi}=C_2r^2\sin^2\theta,								\notag\\
\Gamma^r{}_{tr}=&C_3, 
	\quad  	\Gamma^r{}_{rr}=0, 
	\quad		\Gamma^r{}_{\theta\theta}=-r, 
	\quad		\Gamma^r{}_{\phi\phi}=-r\sin^2\theta,												\notag\\
\Gamma^\theta{}_{t\theta}=&C_3, 
	\quad		\Gamma^\theta{}_{r\theta}=\frac1r,
	\quad		\Gamma^\theta{}_{\phi\phi}=-\cos\theta\sin\theta,										\notag\\
\Gamma^\phi{}_{t\phi}=&C_3, 
	\quad 	\Gamma^\phi{}_{r\phi}=\frac1r, 
	\quad 	\Gamma^\phi{}_{\theta\phi}=\cot\theta.
\end{align}

\begin{enumerate}
\item[(A)] Connection I: $C_1=\gamma$, $C_2=C_3=0$, where $\gamma$ is a function on $t$; or
\item[(B)] Connection II: $C_1=\gamma+\dfrac{\dot\gamma}\gamma$, $C_2=0$, $C_3=\gamma$,
             where $\gamma$ is a nonvanishing function on $t$; or 
\item[(C)] Connection III: $C_1=-\dfrac{\dot\gamma}{\gamma}$, $C_2=\gamma$, $C_3=0$,
             where $\gamma$ is a nonvanishing function on $t$.
\end{enumerate} 
We can then calculate the required tensors in the general setting
\begin{align}
Q_{ttt}=&2C_1,
\quad Q_{tr}{}^r=Q_{t\theta}{}^{\theta}=Q_{t\phi}{}^{\phi}=-2\left(C_3-H\right),
\quad Q^r{}_{rt}=Q^{\theta}{}_{\theta t}=Q^{\phi}{}_{\phi t}=-C_3+\frac{C_2}{a^2}\,,
\end{align}
\begin{align}
L_{ttt}=&-C_1,
\quad L_{tr}{}^{r}=L_{t \theta}{}^{\theta}=L_{t \phi}{}^{\phi}=H-\frac{C_2}{a^2}, 
\quad L^{r}{}_{rt}=L^{\theta}{}_{\theta t}=L^{\phi}{}_{\phi t}=C_3-H\,,
\end{align}
\begin{align}
P_{ttt}=&\frac34\left(-C_3+\frac{C_2}{a^2}\right), 
\quad P_{tr}{}^{r}=P_{t\theta}{}^\theta=P_{t\phi}{}^\phi
				=\frac14\left(4H-3C_3-\frac{C_2}{a^2}\right) 
            , \quad 
P^r{}_{rt}=P^\theta{}_{\theta t}=P^\phi{}_{\phi t}
				=\frac14\left(C_1+C_3-H\right). 
\end{align}
It follows from the above data that the non-metricity scalar is given by 
\begin{align}\label{Q}
Q(t)=3\left(
-2H^2+3C_3H+\frac{C_2}{a^2}H-(C_1+C_3)\frac{C_2}{a^2}+(C_1-C_3)C_3\right).
\end{align}
The Friedmann-like equations corresponding to the field equation (\ref{FE}) are given by
\begin{align}\label{rho}
 \kappa\rho^{\text m}
=&\frac12f+\left(3H^2-\frac12Q\right)f_Q+\frac32\dot Q\left(C_3-\frac{C_2}{a^2}\right)f_{QQ}\,, \\
\label{p}
\kappa p^{\text m}
=&-\frac12f+\left(-2\dot H-3H^2+\frac12Q\right)f_Q
        +\frac12\dot Q\left(-4H+3C_3+\frac{C_2}{a^2}\right)f_{QQ}.
\end{align}
The effective energy density and pressure can also be derived
\begin{align}\label{rho_eff}
\kappa \rho^{\text{eff}}
=&\kappa \rho^{\text m}-\frac12f+\frac{1}{2}Qf_Q-\frac32\dot Q\left(C_3-\frac{C_2}{a^2}\right)f_{QQ}\,,\\
\kappa p^{\text{eff}}
=& \kappa p^{\text m}+\frac12f-\frac{1}{2}Qf_Q
   -\frac12\dot Q\left(-4H+3C_3+\frac{C_2}{a^2}\right)f_{QQ}.
\label{p_eff}
\end{align}

\section{Energy conditions in late-time: cosmographical analysis}
We define the necessary cosmological parameters: the Hubble parameter, $H(t)$, deceleration parameter, $q(t)$ and jerk parameter, $j(t)$ are given by
\begin{align}
    H(t)=&\frac{\dot{a}(t)}{a(t)}, \label{eqn:H}\\
    q(t)=&-\frac{\ddot{a}(t)}{a(t) H(t)^2}, \label{eqn:q}\\
    j(t)=&\frac{\dddot{a}(t)}{a(t) H(t)^3}.
\end{align}
By using the relations above, we have the following
\begin{align}
 \dot{H}=&-H^2(1+q),  \\
 \ddot{H}=&H^3(2+3q+j).
\end{align}
Furthermore, the parameters at the present time are indicated by a $0$ as a subscript e.g., $H_0, j_0, q_0$. We utilise the present day observed values of some cosmological parameters, namely, the Hubble parameter $H_0=69.3 km s^{-1} Mpc^{-1}$, the deceleration parameter $q_0= -0.73$ and the jerk parameter $j_0=2.84$ \cite{capozziello2022a}, to study the late-time accelerated state of the universe, in particular. We consider the present value of the scale factor $a_0=1$. These data provide us with an effective EoS, $\omega^{eff}=\frac{-3H^2-2\dot{H}}{3H^2}=-0.82$ at the present time.

In the following three subsections, we analyse the ECs for all the three classes of connections described in the previous section.
Two of the most important $f(Q)$ models, namely, $f(Q)=Q+\beta Q^2$ and $f(Q)=Q+\beta\sqrt{-Q}$ are used for this purpose. 
The first model is the simplest generalisation of the GR situation ($f(Q)=Q$) and a natural starting choice for almost every modified theories. 
Whereas, the second model is produced in an attempt to mimic a $\Lambda$-cold dark matter ($\Lambda$CDM) evolution in $f(Q)$ theory with the coincident gauge choice for the affine connections of type I with vanishing $\gamma$ in spatially flat FLRW spacetime \cite{cosmology_Q}, later this model was extensively studied in \cite{redshift,signature,lcdm1,siren}. By testing this model against redshift space distortion data, the deviations of evolution of the matter fluctuations were shown to be sufficient to alleviate the present $\sigma_8$ tension \cite{redshift}.
A likelihood analysis was also carried out and a reasonable value of $\beta$ was shown to be $ 2.0331^{+3.8212}_{-1.9596}$. 
On the other hand, in \cite{signature} the authors showed that positive values of $\beta$ suppressed matter power spectrum and lensing effect on the Cosmic Microwave Background radiation (CMB) angular power spectrum, whereas, contrary to the previous study, a negative value was also observed with a completely opposite effect. 
The model was later also discussed in \cite{lcdm1}, where it was shown that the prescribed value of $\beta$ highly impacts the dynamics of the linear matter perturbations and gravitational potentials, a positive value was preferred in this study. 
Very recently, a Bayesian analysis resorting to generated mock catalogs showed that the Laser Interferometer Gravitational-Wave Observatory (LIGO) is not able to distinguish this model from $\Lambda$CDM, while both the Laser Interferometer Space Antenna (LISA) and the The Einstein Telescope (ET) will, with the ET outperforming LISA \cite{siren}. 
However, only the spatially flat FLRW line element with the class of Connection I with vanishing $\gamma$ in Cartesian coordinates (coincident gauge choice) was used in all these past studies.
Due to all these controversies related to the value of $\beta$, it is worthwhile to analyse this model parameter from energy conditions in all possible connections.

\subsection{Connection I: $C_1=\gamma$, $C_2=C_3=0$. }\label{sec2-1}
For the first class of connections, using (\ref{Q})--(\ref{p}), we express the non-metricity scalar $Q$, and the pressure and energy density as 
\begin{align}\label{Q-1}
Q=&-6H^2\,, \\
\label{rho-1}
\kappa \rho^{\text m}=&\frac12f+6H^2f_Q\,, \\
\label{p-1}
\kappa p^{\text m}=&-\frac12f+\left(-2\dot H-6H^2\right)f_Q+24H^2\dot Hf_{QQ}.
\end{align}
These show that the unknown parameter $\gamma(t)$ does not play any role in either of the above terms. So practically speaking, the dynamics of $f(Q)$ theory in the whole class of Connection I can be uniquely determined by (\ref{Q-1})--(\ref{p-1}). This connection type actually reduces to the coincident gauge's representation  if we do a coordinate transformation to a Cartesian one
while $\gamma=0$. In vacuum case, integration of (\ref{rho-1}) straightforward returns $f(Q)\propto \sqrt{-Q}$. However, such a choice forces the Lagrangian into a total derivative, the action becomes a pure surface term. On the other hand, if we try to mimic a $\Lambda$CDM dynamics by comparing $\rho^m$ with $3H^2=\frac Q2$, we can integrate to obtain a model  $f(Q)=Q+\beta \sqrt{-Q}$ \cite{cosmology_Q}. This class (depending on the parameter $\beta$) of models displays a background cosmology equivalent to GR in the connections of  type I. However, it will show a beyond-GR characteristics when we step into the other connection types.

Although trivial, for completeness of the study we briefly disclose the energy condition bounds of the model parameter $\beta$ in Connection I using the two $f(Q)$ models as follows.
\subsubsection{Model 1: $f(Q)=Q+\beta Q^2$}
From (\ref{rho-1})--(\ref{p-1}), we can write the required expressions of ECs as

\begin{align}
\kappa \rho^m=& 14407.5-1.24545\times 10^9 \beta \label{c1wec}\,,\\
\kappa (\rho^m-p^m)=&26221.6-2.04254\times 10^9 \beta \label{c1dec}\,,\\
\kappa (\rho^m+p^m)=&2593.34-4.48362\times 10^8 \beta \label{c1nec}\,, \\
\kappa (\rho^m+3p^m)=&-21034.9+1.14582\times 10^9 \beta\label{c1sec}\,.
\end{align}
It is obvious from the equations (\ref{c1wec})--(\ref{c1nec}) that $\rho^m$, $\rho^m-p^m$, $\rho^m+p^m$, all are non-negative in the range $\beta\le 5.784\times 10^{-6}$, whereas  $\rho^m+3p^m\ge 0$ is satisfied for $\beta\ge 0.000018$.
\subsubsection{Model 2: $f(Q) = Q + \beta\sqrt{-Q}$}
For this model we obtain the following expressions
\begin{align}
\kappa \rho^m=&14407.5\,,\label{c1wecm2}\\
\kappa (\rho^m-p^m)=&26221.6\,, \label{c1decm2}\\
\kappa (\rho^m+p^m)=&2593.34\,,\label{c1necm2}\\
\kappa (\rho^m+3p^m)=&-21034.9\,. \label{c1secm2}
\end{align}
Note that in (\ref{c1wecm2})--(\ref{c1decm2}), all the coefficients of $\beta$ vanish. Indeed, this is quite expected as this particular $f(Q)=Q+\beta\sqrt{-Q}$ model is dynamically equivalent to GR when formulated from connection class I. Consequently, the SEC of ordinary matter is violated.

\subsection{Connection II: $C_1=\gamma+\dfrac{\dot\gamma}\gamma$, $C_2=0$, $C_3=\gamma$.}\label{sec2-2}
This is the first class of connections that provides us novel insight of the $f(Q)$ dynamics in the spatially flat FLRW background, keeping it completely aloof from the $f(T)$ theory. We begin by deriving the energy and pressure equations in this case and then continue to analyse three cases for the unknown parameter $\gamma(t)$ for each of the two $f(Q)$ models, as follows. Let us keep in mind that $\gamma(t)$ cannot be zero for connection class II and III. From (\ref{Q})--(\ref{p}), we can write the non-metricity scalar, the equations of energy density and pressure as
\begin{align}
 Q=&-6H^2+9\gamma H+3\dot\gamma \,,
            \label{Q-2} \\
 \kappa \rho^{\text m}=&\frac12f+\left(6H^2-\frac92\gamma H-\frac32\dot\gamma\right)f_Q
            +\frac92\left(-4H\dot H+3\dot\gamma H+3\gamma\dot H+\ddot\gamma\right)\gamma f_{QQ}           \,,    
            \label{rho-2}\\
 \kappa p^{\text m}=&-\frac12f+\left(-2\dot H-6H^2+\frac92\gamma H+\frac32\dot\gamma \right)f_Q
        +\frac32\left(-4H\dot H+3\dot\gamma H+3\gamma\dot H+\ddot\gamma\right)\left(-4H+3\gamma\right)f_{QQ}\,.
            \label{p-2}
\end{align}
Now, we consider two $f(Q)$ models and for each model, three possible cases of $\gamma(t)$. The expressions required to analyse all the ECs, i.e., $\rho^m,\,\rho^m-p^m,\,\rho^m+p^m,\,\rho^m+3p^m$ are computed using (\ref{Q-2})--(\ref{p-2}), for respective cases and the contour plots are presented.
\subsubsection{Model 1: $f(Q)=Q+\beta Q^2$}
\begin{center}
    {\underline{\bf{Case 1}: \boldsymbol{$\gamma=\gamma_0$}}}
\end{center}
\begin{align}
\kappa \rho^m=&\beta\left(-229511.0\gamma_0^2+3.91787\times 10^7\gamma_0-1.24545\times10^9\right)+14407.5\,,\\
\kappa (\rho^m-p^m)
    =&\beta\left(-389002.0\gamma_0^2+6.54176\times 10^7\gamma_0-2.04254\times10^9\right)+26221.6\,,\\
\kappa (\rho^m+p^m)
    =&\beta\left(-70020.3\gamma_0^2+1.29398\times 10^7\gamma_0-4.48362\times10^8\right)+2593.34\,,\\
\kappa (\rho^m+3p^m)
    =&\beta\left(248961.0\gamma_0^2-3.95381\times 10^7\gamma_0+1.14582\times10^9\right)-21034.9\,.
\end{align}

\begin{figure}[H]
 \begin{minipage}[b]{0.49\textwidth}
   \includegraphics[width=\textwidth]{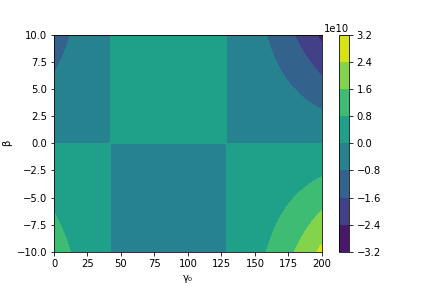}
   \caption{$\kappa \rho^m$}
   \label{fig8}
 \end{minipage}
 \hfill
 \begin{minipage}[b]{0.49\textwidth}
   \includegraphics[width=\textwidth]{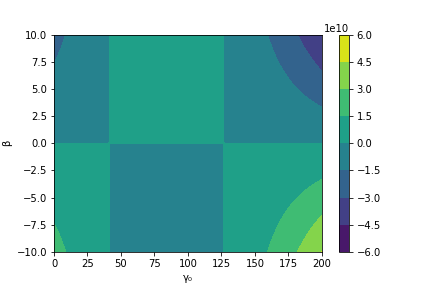}
   \caption{$\kappa (\rho^m-p^m)$}
   \label{fig9}
 \end{minipage}
 \end{figure}
 \begin{figure}[H]
 \begin{minipage}[b]{0.49\textwidth}
   \includegraphics[width=\textwidth]{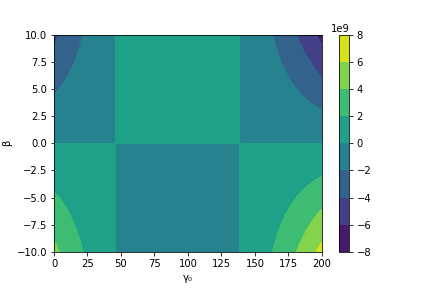}
   \caption{$\kappa (\rho^m+p^m)$}
   \label{fig10}
 \end{minipage}
 \hfill
 \begin{minipage}[b]{0.49\textwidth}
   \includegraphics[width=\textwidth]{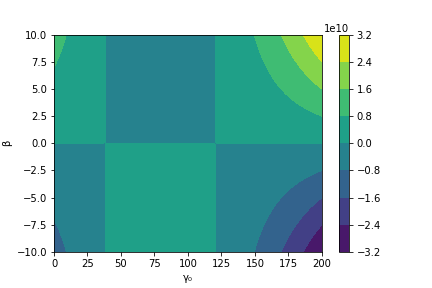}
   \caption{$\kappa (\rho^m+3p^m)$}
   \label{fig11}
 \end{minipage}
\end{figure}
In this case, we observe that WEC, DEC, NEC are satisfied in the range $(46.2<\gamma_0<126.74, \beta\ge0)$, $(\gamma_0<41.43,\beta\le 0)$ and $(\gamma_0>138.6,\beta\le 0)$. SEC is satisfied for ($38.13<\gamma_0<120.67, \beta<0$), ($0<\gamma_0<38.13, \beta>0$) and ($\gamma_0>120.67, \beta>0$). 

\begin{center}
    {\underline{\bf{Case 2}: \boldsymbol{$\gamma=\gamma_1 a(t)$}}}
\end{center}

\begin{align}
    \kappa \rho^m=&\beta\left(-219570.0\gamma_1^2+5.11599\times10^7\gamma_1-1.24545\times10^9\right)+14407.5\,,\\
    \kappa (\rho^m-p^m)=&\beta\left(-691559.0\gamma_1^2+1.03199\times10^8\gamma_1-2.04254\times 10^9\right)+26221.6\,,\\
    \kappa (\rho^m+p^m)=&\beta\left(252419.0\gamma_1^2-878625.0\gamma_1-4.48362\times10^8\right)+2593.34\,,\\
    \kappa (\rho^m+3p^m)=&\beta\left(1.1964\times 10^6\gamma_1^2-1.04956\times10^8\gamma_1+1.14582\times10^9\right)-21034.9\,.
\end{align}
\begin{figure}[H]
 \begin{minipage}[b]{0.49\textwidth}
   \includegraphics[width=\textwidth]{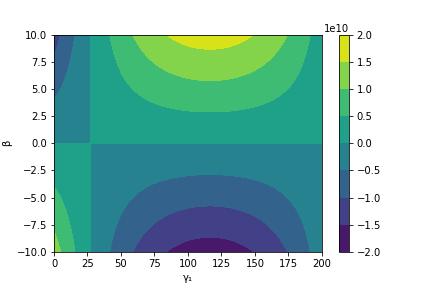}
   \caption{$\kappa \rho^m$}
   \label{fig12}
 \end{minipage}
 \hfill
 \begin{minipage}[b]{0.49\textwidth}
   \includegraphics[width=\textwidth]{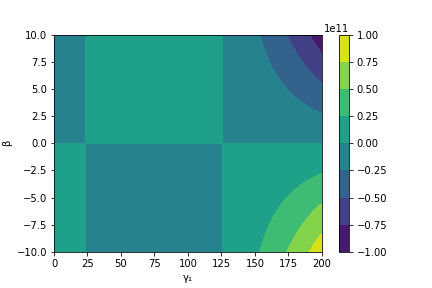}
   \caption{$\kappa (\rho^m-p^m)$}
   \label{fig13}
 \end{minipage}
  \end{figure}
 \begin{figure}[H]
 \begin{minipage}[b]{0.49\textwidth}
   \includegraphics[width=\textwidth]{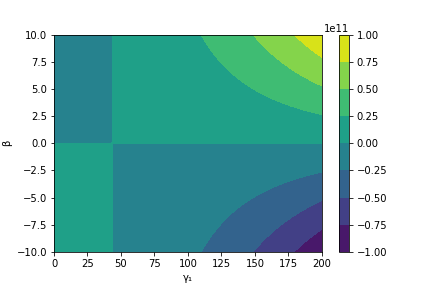}
   \caption{$\kappa (\rho^m+p^m)$}
   \label{fig14}
 \end{minipage}
 \hfill
 \begin{minipage}[b]{0.49\textwidth}
   \includegraphics[width=\textwidth]{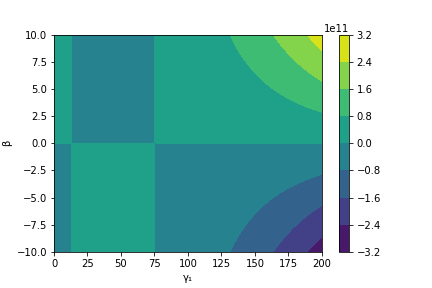}
   \caption{$\kappa (\rho^m+3p^m)$}
   \label{fig15}
 \end{minipage}
\end{figure}
In this case, we observe that WEC, DEC, NEC are satisfied in the range $(43.92<\gamma_1<125.74, \beta\ge 0)$ and  $(\gamma_0<23.49, \beta\le 0)$. SEC is satisfied for ($12.78<\gamma_1<74.95,\beta< 0$), ($\gamma_1<12.78, \beta>0$) and ($\gamma_1>74.95, \beta>0$).
\begin{center}
    {\underline{\bf{Case 3}: \boldsymbol{$\gamma=\gamma_2 H(t)$}}}
\end{center}
\begin{align}
    \kappa \rho^m=&\beta\left(-5.59716\times 10^8\gamma_2^2+2.4909\times 10^9\gamma_2-1.24545\times 10^9\right)+14407.5\,,\\
    \kappa (\rho^m-p^m)=&\beta\left(-1.54704\times 10^9\gamma_2^2+4.61451\times 10^9\gamma_2-2.04254\times 10^9\right)+26221.6\,,\\
    \kappa (\rho^m+p^m)=&\beta\left(4.27605\times 10^8\gamma_2^2+3.67297\times 10^8\gamma_2-4.48362\times 10^8\right)+2593.34\,,\\
    \kappa (\rho^m+3p^m)=&\beta\left(2.40225\times 10^9\gamma_2^2-3.87991\times 10^9\gamma_2+1.14582\times 10^9\right)-21034.9\,.
\end{align}
\begin{figure}[H]
 \begin{minipage}[b]{0.49\textwidth}
   \includegraphics[width=\textwidth]{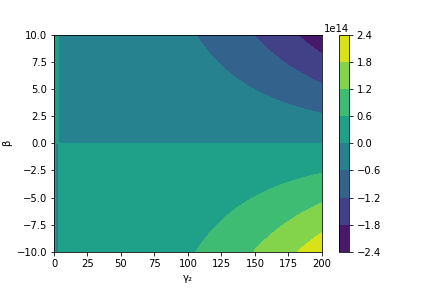}
   \caption{$\kappa \rho^m$}
   \label{fig16}
 \end{minipage}
 \hfill
 \begin{minipage}[b]{0.49\textwidth}
   \includegraphics[width=\textwidth]{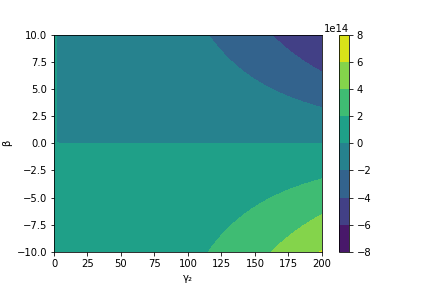}
   \caption{$\kappa (\rho^m-p^m)$}
   \label{fig17}
 \end{minipage}
  \end{figure}
 \begin{figure}[H]
 \begin{minipage}[b]{0.49\textwidth}
   \includegraphics[width=\textwidth]{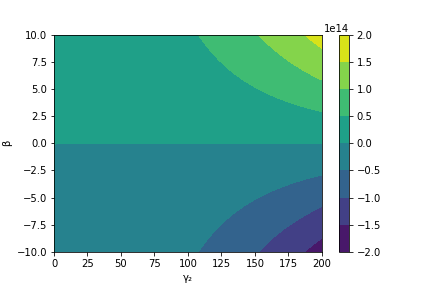}
   \caption{$\kappa (\rho^m+p^m)$}
   \label{fig18}
 \end{minipage}
 \hfill
 \begin{minipage}[b]{0.49\textwidth}
   \includegraphics[width=\textwidth]{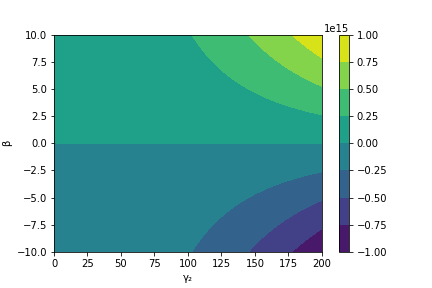}
   \caption{$\kappa (\rho^m+3p^m)$}
   \label{fig19}
 \end{minipage}
\end{figure}
In this case, we observe that NEC and WEC are satisfied for the range $(\gamma_2<0.54, \beta\ge 0)$ and $(\gamma_2>0.57, \beta\le 0)$. Whereas, DEC and SEC are satisfied for $(\gamma_2>0, \beta>0)$.
\begin{center}
    \underline{{\bf{Case 4}: {\boldsymbol{$\gamma=\frac{\gamma_3}{a^{3}}$}}}}
\end{center}
\begin{align}
    \kappa \rho^m=&\beta  \left(3.23494\times 10^6 \gamma _3-1.24545\times 10^9\right)+14407.5\,,\\
    \kappa (\rho^m-p^m)=&26221.6-2.04254\times 10^9 \beta\,,\\
   \kappa (\rho^m+p^m)=&\beta  \left(6.46988\times 10^6 \gamma _3-4.48362\times 10^8\right)+2593.34\,,\\
    \kappa (\rho^m+3p^m)=&\beta  \left(1.29398\times 10^7 \gamma _3+1.14582\times 10^9\right)-21034.9\,.\label{ec566}
\end{align}
\begin{figure}[H]
 \begin{center} 
 \begin{minipage}[b]{0.65\textwidth}
   \includegraphics[width=\textwidth]{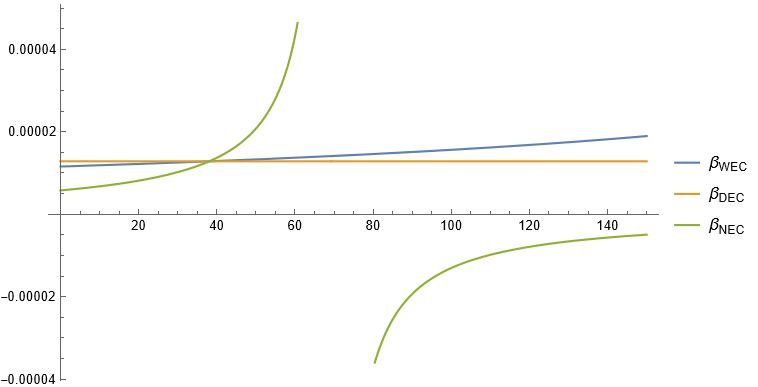}
   \caption{$\beta_{\text{WEC}}, \beta_{\text{DEC}}, \beta_{\text{NEC}}$ vs $\gamma_3$ for vanishing $\rho^m,\, \rho^m-p^m$ and $\rho^m+p^m$}
   \label{fig200}
 \end{minipage}
 \end{center}
\end{figure}
From FIG (\ref{fig200}), we observe that for $\gamma_3<38.08$, WEC, DEC and NEC are satisfied when $\beta$ lies in the region above the orange line. For $38.08<\gamma_3<69.3$, $\beta$ must lie in the region above the green line, whereas for $\gamma>69.3$, $\beta$ must lie in the region above the blue line, for a valid NEC,DEC and WEC. SEC is satisfied for any $\gamma_3\ge0$, as long as $\beta$ is reasonably large, as can be clearly seen in (\ref{ec566}), for example, if $\gamma_3=0$, one needs $\beta>0.00002$.

\subsubsection{Model 2: $f(Q) = Q + \beta\sqrt{-Q}$}
\begin{center}
    {\underline{\bf{Case 1}: \boldsymbol{$\gamma=\gamma_0$}}}
\end{center}
\begin{align}
    \kappa \rho^m=&-\frac{0.069354\beta\left(-94.0744\gamma_0^2+4533.4\gamma_0\right)}{\left(46.2-\gamma_0\right)^{\frac{3}{2}}}+14407.5\,,\label{sqrtconn2wec}\\
    \kappa (\rho^m-p^m)=&-\frac{0.069354\beta\left(-180.047\gamma_0^2+8692.47\gamma_0\right)}{\left(46.2-\gamma_0\right)^{\frac{3}{2}}}+26221.6\,,\\
    \kappa (\rho^m+p^m)=&2593.34-\frac{0.56191\beta\gamma_0}{\left(46.2-\gamma_0\right)^{\frac{1}{2}}}\,,\label{sqrtconn2nec}\\
    \kappa (\rho^m+3p^m)=&-\frac{0.069354\beta\left(163.842\gamma_0^2-7943.84\gamma_0\right)}{\left(46.2-\gamma_0\right)^{\frac{3}{2}}}-21034.9\,.\label{sqrtconn2sec}
\end{align}
\begin{figure}[H]
 \begin{minipage}[b]{0.49\textwidth}
   \includegraphics[width=\textwidth]{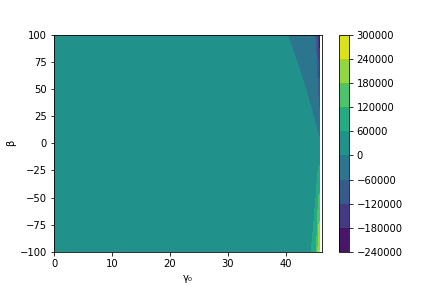}
   \caption{$\kappa \rho^m$}
   \label{fig20}
 \end{minipage}
 \hfill
 \begin{minipage}[b]{0.49\textwidth}
   \includegraphics[width=\textwidth]{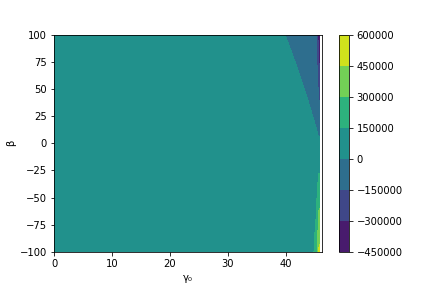}
   \caption{$\kappa (\rho^m-p^m)$}
   \label{fig21}
 \end{minipage}
 \hfill
 \begin{minipage}[b]{0.49\textwidth}
   \includegraphics[width=\textwidth]{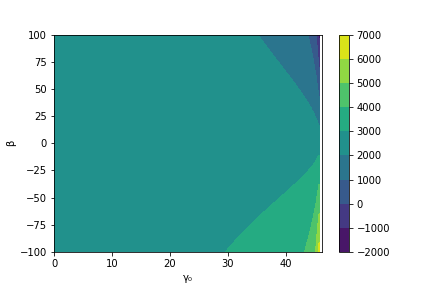}
   \caption{$\kappa (\rho^m+p^m)$}
   \label{fig22}
 \end{minipage}
 \hfill
 \begin{minipage}[b]{0.49\textwidth}
   \includegraphics[width=\textwidth]{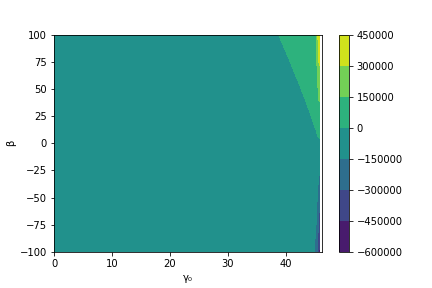}
   \caption{$\kappa (\rho^m+3p^m)$}
   \label{fig23}
 \end{minipage}
 \end{figure}
 \begin{figure}[H]
 \begin{center}
  \begin{minipage}[b]{0.65\textwidth}
   \includegraphics[width=\textwidth]{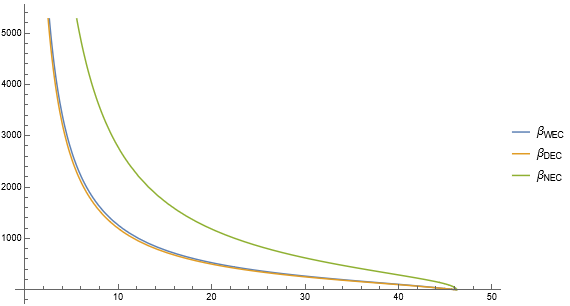}
   \caption{$\beta_{\text{WEC}}, \beta_{\text{DEC}}, \beta_{\text{NEC}}$ vs $\gamma_0$ for vanishing $\rho^m,\, \rho^m-p^m$ and $\rho^m+p^m$}
   \label{fig24}
  \end{minipage}
  \end{center}
\end{figure}
Clearly, $\gamma_0<46.2$ in this case. The ranges for all the ECs are clearly depicted in FIG. \ref{fig20}--\ref{fig23} above. From (\ref{sqrtconn2wec})--(\ref{sqrtconn2nec}), we observe that NEC, WEC, DEC are satisfied for a dynamic range of 
$$\beta< \frac{378083.46\left(46.2-\gamma_0\right)^{\frac{3}{2}}}{-180.047\gamma_0^2+8692.47\gamma_0}.$$
FIG.~\ref{fig24} displays the maximum possible value of $\beta$ for any permitted range of $\gamma_0$, so that NEC, WEC and DEC are satisfied. Whereas, (\ref{sqrtconn2sec}) shows that SEC is satisfied for 
$$\beta\le-\frac{303297.57\left(46.2-\gamma_0\right)^{\frac{3}{2}}}{163.842\gamma_0^2-7943.84\gamma_0}.$$ Take note of the change of inequality when the denominator is negative for certain values of $\gamma_0$.
\begin{center}
 {\underline{\bf{Case 2}: \boldsymbol{$\gamma=\gamma_1 a(t)$}}}
\end{center}
\begin{align}
    \kappa \rho^m=&-\frac{0.060063\beta\left(4439.82\gamma_1-109.078\gamma_1^2\right)}{\left(34.65-\gamma_1\right)^{\frac{3}{2}}}+14407.5\,,\\
    \kappa (\rho^m-p^m)=&-\frac{0.060063\beta\left(10078.8\gamma_1-240.062\gamma_1^2\right)}{\left(34.65-\gamma_1\right)^{\frac{3}{2}}}+26221.6\,,\\
    \kappa (\rho^m+p^m)=&-\frac{0.060063\beta\left(21.9057\gamma_1^2-1199.2\gamma_1\right)}{\left(34.65-\gamma_1\right)^{\frac{3}{2}}}+2593.34\,,\\
    \kappa (\rho^m+3p^m)=&-\frac{0.060063\beta\left(283.874\gamma_1^2-12477.2\gamma_1\right)}{\left(34.65-\gamma_1\right)^{\frac{3}{2}}}-21034.9\,.
\end{align}
The following are the plots respectively
\begin{figure}[H]
 \begin{minipage}[b]{0.49\textwidth}
   \includegraphics[width=\textwidth]{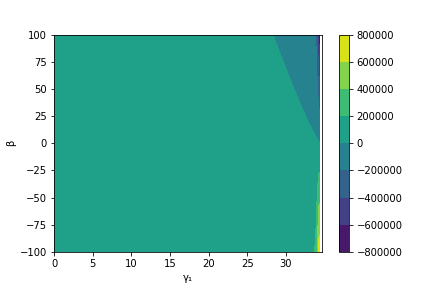}
   \caption{$\kappa \rho^m$}
   \label{fig25}
 \end{minipage}
 \hfill
 \begin{minipage}[b]{0.49\textwidth}
   \includegraphics[width=\textwidth]{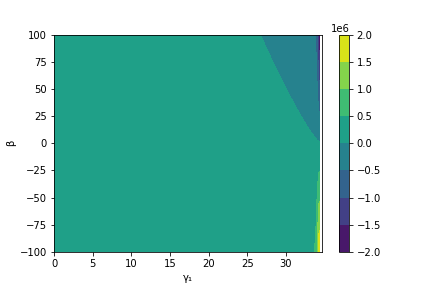}
   \caption{$\kappa (\rho^m-p^m)$}
   \label{fig26}
 \end{minipage}
 \end{figure}
 \begin{figure}[H]
 \begin{minipage}[b]{0.49\textwidth}
   \includegraphics[width=\textwidth]{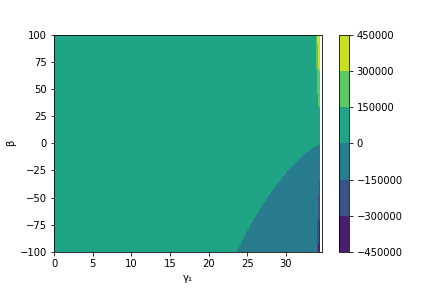}
   \caption{$\kappa (\rho^m+p^m)$}
   \label{fig27}
 \end{minipage}
 \hfill
 \begin{minipage}[b]{0.49\textwidth}
   \includegraphics[width=\textwidth]{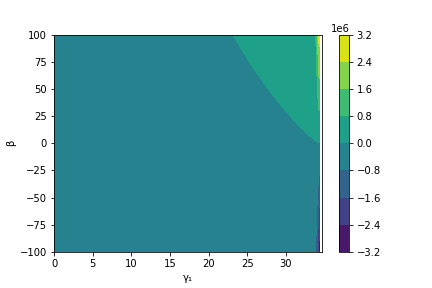}
   \caption{$\kappa (\rho^m+3p^m)$}
   \label{fig28}
 \end{minipage}
 \end{figure}
 
 \begin{figure}[H]
 \begin{center} 
 \begin{minipage}[b]{0.65\textwidth}
   \includegraphics[width=\textwidth]{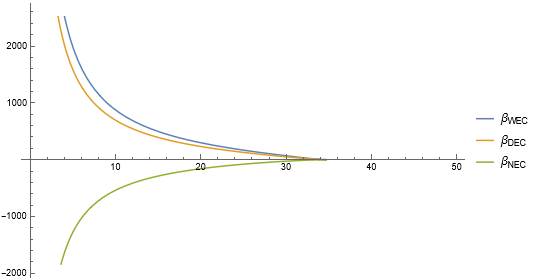}
   \caption{$\beta_{\text{WEC}}, \beta_{\text{DEC}}, \beta_{\text{NEC}}$ vs $\gamma_1$ for vanishing $\rho^m,\, \rho^m-p^m$ and $\rho^m+p^m$}
   \label{fig29}
 \end{minipage}
 \end{center}
\end{figure}
In this case $\gamma_0<34.65$. From figure (\ref{fig29}), we observe that NEC, WEC, DEC are satisfied in the range 
\[\frac{43176\left(34.65-\gamma_1\right)^{\frac{3}{2}}}{21.9057\gamma_1^2-1199.2\gamma_1}<\beta<\frac{436568.2\left(34.65-\gamma_1\right)^{\frac{3}{2}}}{10078.8\gamma_1-240.062\gamma_1^2}. \]
Whereas, SEC is satisfied for 
\[\beta\le-\frac{350213.9\left(34.65-\gamma_1\right)^{\frac{3}{2}}}{283.874\gamma_1^2-12477.2\gamma_1}.\] Take note of the change of the inequality when the denominator is negative for certain values of $\gamma_1$.
\begin{center}
 {\underline{\bf{Case 3}: \boldsymbol{$\gamma=\gamma_2 H(t)$}}}
\end{center}
\begin{align}
    \kappa \rho^m=&-\frac{0.008733\beta\left(-5284.85\gamma_2^2+4570.42\gamma_2\right)}{\left(0.732601-\gamma_2\right)^{\frac{3}{2}}}+14407.5\,,\\
    \kappa (\rho^m-p^m)=&-\frac{0.008733\beta\left(-11354.3\gamma_2^2+9589.85\gamma_2\right)}{\left(0.732601-\gamma_2\right)^{\frac{3}{2}}}+26221.6\,,\\
    \kappa (\rho^m+p^m)=&-\frac{0.008733\beta\left(784.588\gamma_2^2-449.018\gamma_2\right)}{\left(0.732601-\gamma_2\right)^{\frac{3}{2}}}+2593.34\,,\\
    \kappa (\rho^m+3p^m)=&-\frac{0.008733\beta\left(12923.5\gamma_2^2-10487.9\gamma_2\right)}{\left(0.732601-\gamma_2\right)^{\frac{3}{2}}}-21034.9\,.
\end{align}
\begin{figure}[H]
 \begin{minipage}[b]{0.49\textwidth}
   \includegraphics[width=\textwidth]{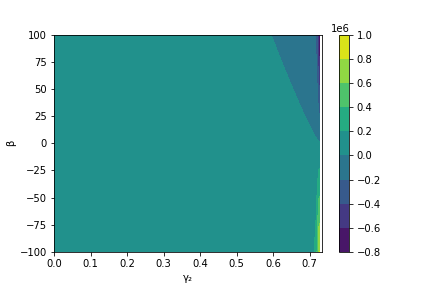}
   \caption{$\kappa \rho^m$}
   \label{fig30}
 \end{minipage}
 \hfill
 \begin{minipage}[b]{0.49\textwidth}
   \includegraphics[width=\textwidth]{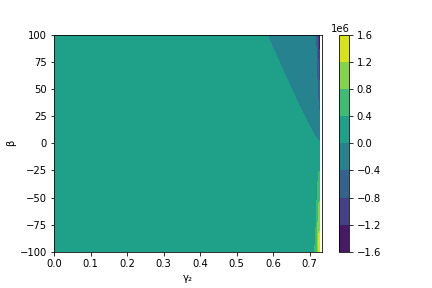}
   \caption{$\kappa (\rho^m-p^m)$}
   \label{fig31}
 \end{minipage}
 \end{figure}
 \begin{figure}[H]
 \begin{minipage}[b]{0.49\textwidth}
   \includegraphics[width=\textwidth]{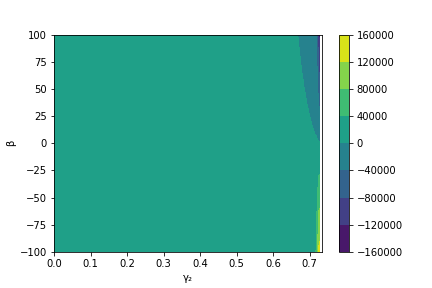}
   \caption{$\kappa (\rho^m+p^m)$}
   \label{fig32}
 \end{minipage}
 \hfill
 \begin{minipage}[b]{0.49\textwidth}
   \includegraphics[width=\textwidth]{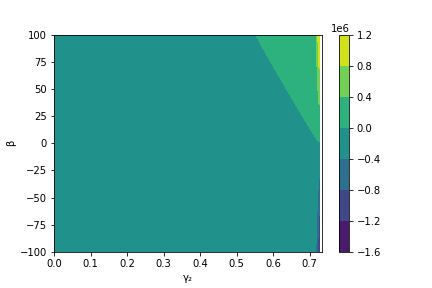}
   \caption{$\kappa (\rho^m+3p^m)$}
   \label{fig33}
 \end{minipage}
 \end{figure}
 \begin{figure}[h]
 \begin{center}
 \begin{minipage}[b]{0.6\textwidth}
   \includegraphics[width=\textwidth]{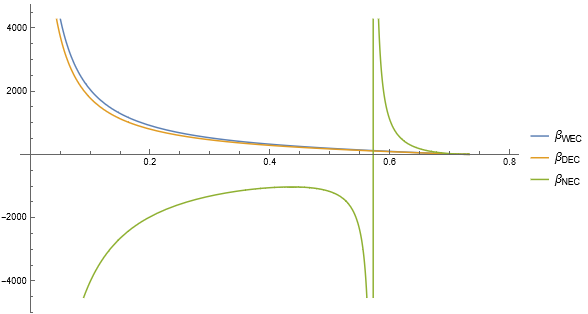}
   \caption{$\beta_{\text{WEC}}, \beta_{\text{DEC}}, \beta_{\text{NEC}}$ vs $\gamma_2$ for vanishing $\rho^m,\, \rho^m-p^m$ and $\rho^m+p^m$}
   \label{fig34}
 \end{minipage}
 \end{center}
\end{figure}

In this case $\gamma_2<0.733$. From figure \ref{fig34}, we observe that NEC, WEC, DEC are satisfied in the range  
\[\frac{296943.66\left(0.732601-\gamma_2\right)^{\frac{3}{2}}}{784.588\gamma_2^2-449.018\gamma_2}<\beta<\frac{3002436.18\left(0.732601-\gamma_2\right)^{\frac{3}{2}}}{-11354.3\gamma_2^2+9589.85\gamma_2}.\] SEC is satisfied for 
\[\beta\le-\frac{2408668.27\left(0.732601-\gamma_2\right)^{\frac{3}{2}}}{12923.5\gamma_2^2-10487.9\gamma_2}.\] Take note of the change of the inequality when the denominator is negative for certain values of $\gamma_2$.
\begin{center}
    \underline{\bf{Case 4}: \boldsymbol{$\gamma=\frac{\gamma_3}{a^{3}}$}}
\end{center}

\begin{align}
    \kappa \rho^m=&14407.5-0.0826703 \beta  \gamma _3\,,\\
    \kappa (\rho^m-p^m)=&26221.6\,,\\
    \kappa (\rho^m+p^m)=&2593.34-0.165341 \beta  \gamma _3\,,\\
    \kappa (\rho^m+3p^m)=&-0.330681 \beta  \gamma _3-21034.9\,.
\end{align}
\begin{figure}[H]
 \begin{center} 
 \begin{minipage}[b]{0.65\textwidth}
   \includegraphics[width=\textwidth]{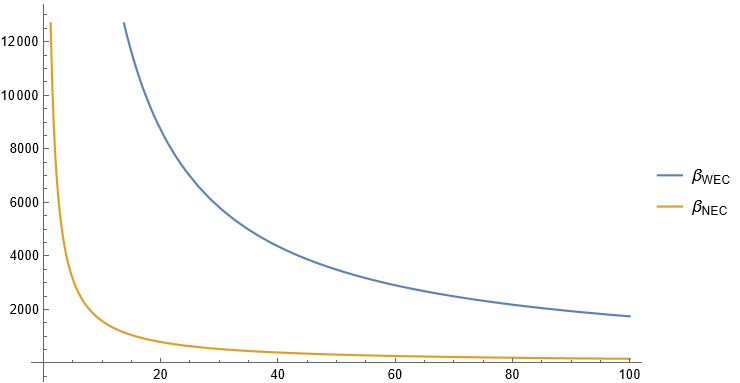}
   \caption{$\beta_{\text{WEC}}, \beta_{\text{NEC}}$ vs $\gamma_3$ for vanishing $\rho^m$ and $\rho^m+p^m$}
   \label{fig203}
 \end{minipage}
 \end{center}
\end{figure}
From FIG (\ref{fig203}), we observe that WEC and NEC are satisfied for $\beta\gamma_3<15684.8$, while DEC remains constant irrespective of $\gamma_3$ and $\beta$. Furthermore, SEC is satisfied for $\beta\gamma_3<-63610.9$.

\subsection{Connection III: $C_1=-\dfrac{\dot\gamma}\gamma$, $C_2=\gamma$, $C_3=0$. }\label{sec2-3}
In a similar manner, from (\ref{Q})--(\ref{p}) we obtain the following expressions  
\begin{align}
Q=&-6H^2+3\frac{\gamma H+\dot\gamma}{a^2}\,,
            \label{Q-3} \\
\kappa \rho^m=&\frac12f+\left(6H^2-\frac32\frac{\gamma H+\dot\gamma}{a^2} \right)f_Q
            -\frac92\left(-4H\dot H+\frac{\gamma\dot H-\dot\gamma H-2\gamma H^2+\ddot\gamma}{a^2}\right)
            \frac\gamma{a^2} f_{QQ}  \,,             
            \label{rho-3}\\
\kappa p^m=&-\frac12f+\left(-2\dot H-6H^2+\frac32\frac{\gamma H+\dot\gamma}{a^2}\right)f_Q
        +\frac32\left(-4H\dot H+\frac{\gamma\dot H-\dot\gamma H-2\gamma H^2+\ddot\gamma}{a^2}\right)\left(-4H+\frac\gamma{a^2}\right)f_{QQ}\,
            \label{p-3}\,.
\end{align}
\subsubsection{Model 1: $f(Q) = Q + \beta Q^2$}
Using equations (\ref{rho-3})--(\ref{p-3}), we derive the required linear combinations of energy and pressure for the respective cases.

\begin{center}
    {\underline{Case 1}: \boldsymbol{$\gamma=\gamma_0$}}
\end{center}
\begin{align}
    \kappa \rho^m=&\beta\left(76503.7\gamma_0^2+8.74631\times10^6\gamma_0-1.24545\times 10^9\right)+14407.5\,,\\
    \kappa (\rho^m-p^m)=&\beta\left(87597.4\gamma_0^2+9.50513\times10^6\gamma_0-2.04254\times 10^9\right)+26221.6\,,\\
    \kappa (\rho^m+p^m)=&\beta\left(65409.9\gamma_0^2+7.9875\times10^6\gamma_0-4.48362\times 10^8\right)+2593.34\,,\\
    \kappa (\rho^m+3p^m)=&\beta\left(43222.4\gamma_0^2+6.46988\times10^6\gamma_0+1.14582\times 10^9\right)-21034.9\,.
\end{align}
\begin{figure}[H]
 \begin{minipage}[b]{0.49\textwidth}
   \includegraphics[width=\textwidth]{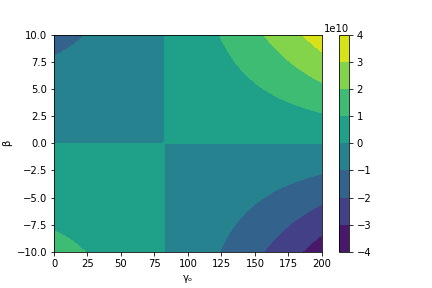}
   \caption{$\kappa \rho^m$}
   \label{fig35}
 \end{minipage}
 \hfill
 \begin{minipage}[b]{0.49\textwidth}
   \includegraphics[width=\textwidth]{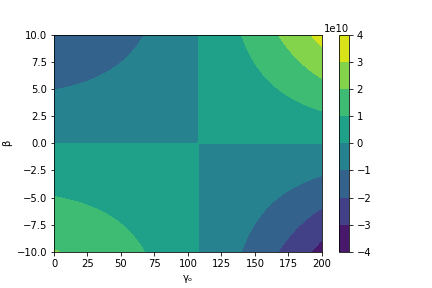}
   \caption{$\kappa (\rho^m-p^m)$}
   \label{fig36}
 \end{minipage}
 \hfill
 \begin{minipage}[b]{0.49\textwidth}
   \includegraphics[width=\textwidth]{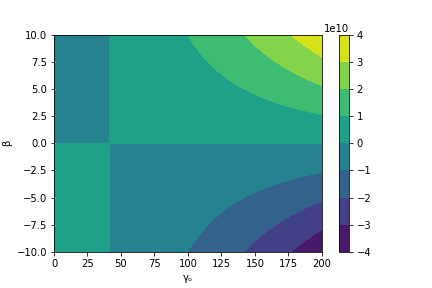}
   \caption{$\kappa (\rho^m+p^m)$}
   \label{fig37}
 \end{minipage}
 \hfill
 \begin{minipage}[b]{0.49\textwidth}
   \includegraphics[width=\textwidth]{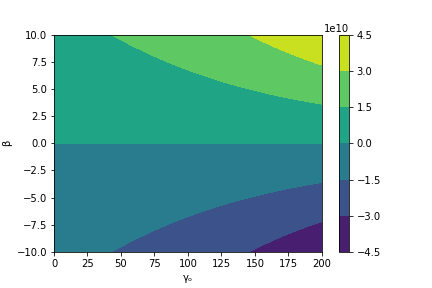}
   \caption{$\kappa (\rho^m+3p^m)$}
   \label{fig38}
 \end{minipage}
\end{figure}

In this case, we observe that NEC, WEC, DEC are satisfied in the range $(\gamma_0>107.80, \beta\ge0)$ and ($\gamma_0<41.82, \beta\le 0$). Whereas, SEC is satisfied for $\beta>0$ for any positive $\gamma_0$. 

\begin{center}
 {\underline{Case 2}: \boldsymbol{$\gamma=\gamma_1 a(t)$}}
\end{center}
\begin{align}
    \kappa \rho^m=&\beta\left(23340.1 \gamma_1^2+2.07276\times 10^7\gamma_1-1.24545\times 10^9\right)+14407.5\,,\\
    \kappa (\rho^m-p^m)=&\beta\left(-26509.7\gamma_1^2+3.1311\times 10^7\gamma_1-2.04254\times 10^9\right)+26221.6\,,\\
    \kappa (\rho^m+p^m)=&\beta\left(73189.9\gamma_1^2+1.01441\times 10^7\gamma_1-4.48362\times 10^8\right)+2593.34\,,\\
    \kappa (\rho^m+3p^m)=&\beta\left(172890.0\gamma_1^2-1.10228\times 10^7\gamma_1+1.14582\times 10^9\right)-21034.9\,.
\end{align}
\begin{figure}[H]
 \begin{minipage}[b]{0.49\textwidth}
   \includegraphics[width=\textwidth]{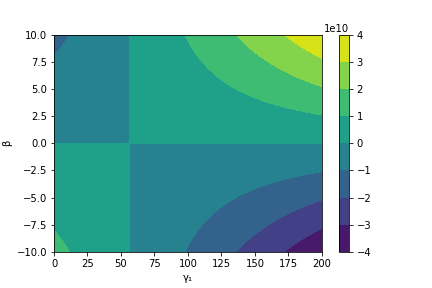}
   \caption{$\kappa \rho^m$}
   \label{fig39}
 \end{minipage}
 \hfill
 \begin{minipage}[b]{0.49\textwidth}
   \includegraphics[width=\textwidth]{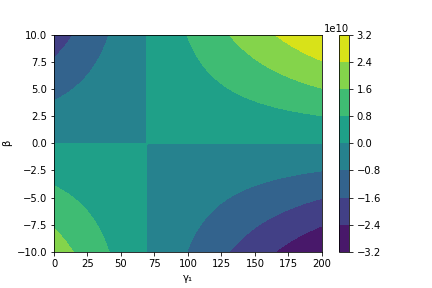}
   \caption{$\kappa (\rho^m-p^m)$}
   \label{fig40}
 \end{minipage}
 \hfill
 \begin{minipage}[b]{0.49\textwidth}
   \includegraphics[width=\textwidth]{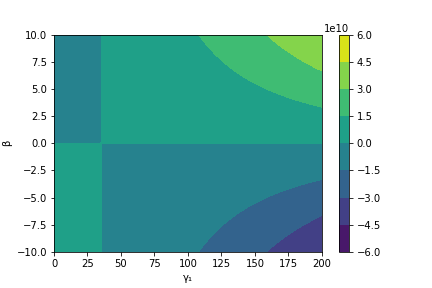}
   \caption{$\kappa (\rho^m+p^m)$}
   \label{fig41}
 \end{minipage}
 \hfill
 \begin{minipage}[b]{0.49\textwidth}
   \includegraphics[width=\textwidth]{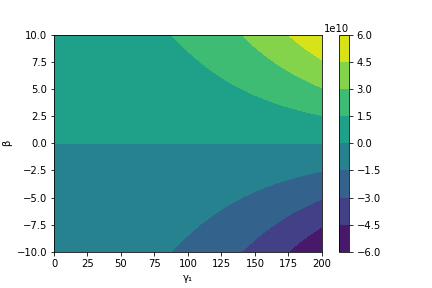}
   \caption{$\kappa (\rho^m+3p^m)$}
   \label{fig42}
 \end{minipage}
\end{figure}
In this case, we observe that NEC, WEC, DEC are satisfied in the range $(69.3<\gamma_1<1111.81, \beta\ge 0)$ and  $(\gamma_1<35.24, \beta<0)$. SEC is satisfied for $\beta>0$ for any positive $\gamma_1$.

\begin{center}
    {\underline{Case 3}: \boldsymbol{$\gamma=\gamma_2 H(t)$}}
\end{center}
\begin{align}
    \kappa \rho^m=&\beta\left(-1.90232\times 10^8\gamma_2^2+3.81938\times 10^8 \gamma_2-1.24545\times 10^9\right)+14407.5\,,\\
    \kappa (\rho^m-p^m)=&\beta\left(-2.90515\times 10^8\gamma_2^2+1.03868\times 10^9\gamma_2-2.04254\times 10^9\right)+26221.6\,,\\
    \kappa (\rho^m+p^m)=&\beta\left(-8.99492\times 10^7\gamma_2^2-2.74802\times 10^8\gamma_2-4.48362\times 10^8\right)+2593.34\,,\\
    \kappa (\rho^m+3p^m)=&\beta\left(1.10617\times 10^8\gamma_2^2-1.58828\times 10^9\gamma_2+1.14582\times 10^9\right)-21034.9\,.
\end{align}
\begin{figure}[H]
 \begin{minipage}[b]{0.49\textwidth}
   \includegraphics[width=\textwidth]{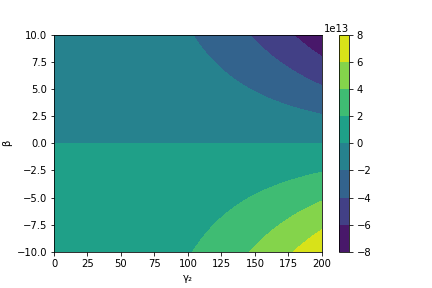}
   \caption{$\kappa \rho^m$}
   \label{fig43}
 \end{minipage}
 \hfill
 \begin{minipage}[b]{0.49\textwidth}
   \includegraphics[width=\textwidth]{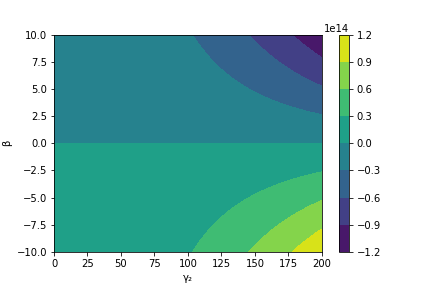}
   \caption{$\kappa (\rho^m-p^m)$}
   \label{fig44}
 \end{minipage}
 \hfill
 \begin{minipage}[b]{0.49\textwidth}
   \includegraphics[width=\textwidth]{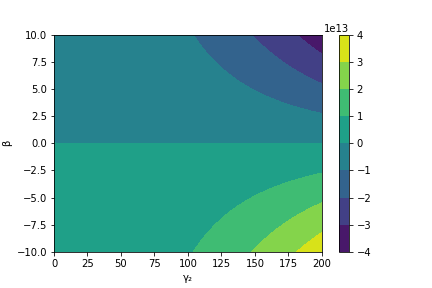}
   \caption{$\kappa (\rho^m+p^m)$}
   \label{fig45}
 \end{minipage}
 \hfill
 \begin{minipage}[b]{0.49\textwidth}
   \includegraphics[width=\textwidth]{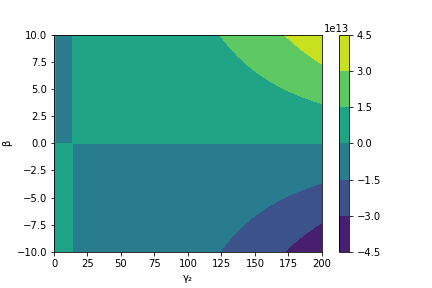}
   \caption{$\kappa (\rho^m+3p^m)$}
   \label{fig46}
 \end{minipage}
\end{figure}
In this case, we observe that NEC, WEC, DEC are satisfied for $\beta\le 0$ and for any positive $\gamma_2$. Whereas, SEC is satisfied for $(\beta<0, \gamma_2<13.597)$ and $(\beta>0, \gamma_2>13.597)$.
\begin{center}
    \underline{\bf{Case 4}: \boldsymbol{$\gamma=\frac{\gamma_4}{a}$}}
\end{center}

\begin{align}
    \kappa \rho^m=&\beta \left(-3.23494\times 10^6 \gamma _4-1.24545\times 10^9\right)+14407.5\,,\\
    \kappa (\rho^m-p^m)=&\beta \left(-4.31325\times 10^6 \gamma _4-2.04254\times 10^9\right)+26221.6\,,\\
    \kappa (\rho^m+p^m)=&\beta  \left(-2.15663\times 10^6 \gamma _4-4.48362\times 10^8\right)+2593.34\,,\\
    \kappa (\rho^m+3p^m)=&1.14582\times 10^9 \beta -21034.9\,.
\end{align}
\begin{figure}[H]
 \begin{center} 
 \begin{minipage}[b]{0.65\textwidth}
   \includegraphics[width=\textwidth]{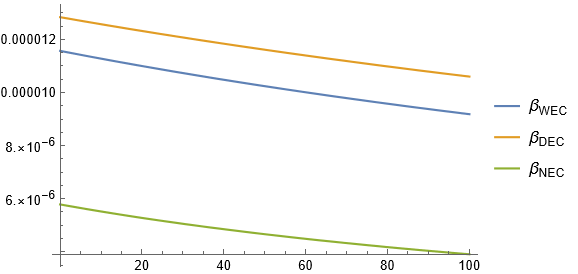}
   \caption{$\beta_{\text{WEC}}, \beta_{\text{DEC}}, \beta_{\text{NEC}}$ vs $\gamma_4$ for vanishing $\rho^m,\, \rho^m-p^m$ and $\rho^m+p^m$}
   \label{fig201}
 \end{minipage}
 \end{center}
\end{figure}
From FIG (\ref{fig201}), we observe that NEC, DEC and WEC are satisfied when $\beta$ lies in the region below the green line, for any reasonable choice of $\gamma$. Whereas, SEC is satisfied for $\beta>1.8358 \times 10^{-5}$ for any $\gamma_4>0$.
\subsubsection{Model 2: $f(Q) = Q + \beta\sqrt{-Q}$}
\begin{center}
    {\underline{\bf{Case 1}: \boldsymbol{$\gamma=\gamma_0$}}}
\end{center}
\begin{align}
    \kappa \rho^m=&\frac{1.8023\beta\left(-0.27\gamma_0^2-202.356\gamma_0\right)}{\left(138.6-\gamma_0\right)^{\frac{3}{2}}}+14407.5\,,\\
    \kappa (\rho^m-p^m)=&\frac{2.4031\beta\left(0.73\gamma_0^2-221.067\gamma _0\right)}{\left(138.6-\gamma_0\right)^{\frac{3}{2}}}+26221.6\,,\\
    \kappa (\rho^m+p^m)=&\frac{1.2016\beta\left(-2.27\gamma_0^2-164.934\gamma_0\right)}{\left(138.6-\gamma_0\right)^{\frac{3}{2}}}+2593.34\,,\\
    \kappa (\rho^m+3p^m)=&\frac{7.20937\beta\left(18.711\gamma _0-\gamma_0^2\right)}{\left(138.6-\gamma_0\right)^{\frac{3}{2}}}-21034.9\,.
\end{align}
\begin{figure}[H]
 \begin{minipage}[b]{0.49\textwidth}
   \includegraphics[width=\textwidth]{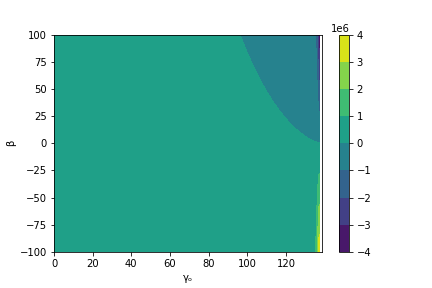}
   \caption{$\kappa \rho^m$}
   \label{fig47}
 \end{minipage}
 \hfill
 \begin{minipage}[b]{0.49\textwidth}
   \includegraphics[width=\textwidth]{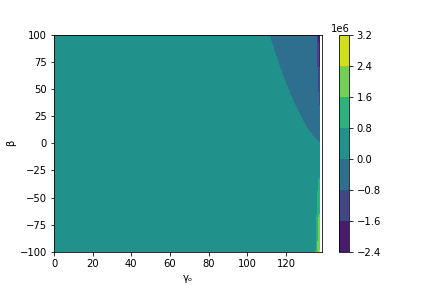}
   \caption{$\kappa (\rho^m-p^m)$}
   \label{fig48}
 \end{minipage}
 \end{figure}
 \begin{figure}[H]
 \begin{minipage}[b]{0.49\textwidth}
   \includegraphics[width=\textwidth]{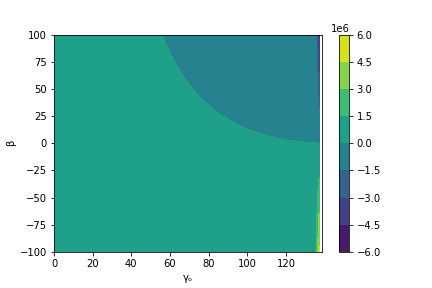}
   \caption{$\kappa (\rho^m+p^m)$}
   \label{fig49}
 \end{minipage}
 \hfill
 \begin{minipage}[b]{0.49\textwidth}
   \includegraphics[width=\textwidth]{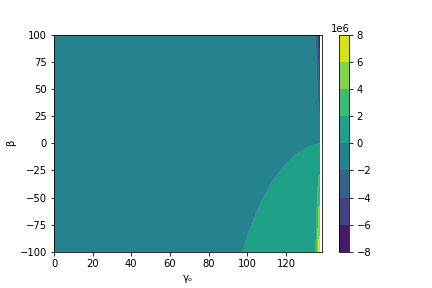}
   \caption{$\kappa (\rho^m+3p^m)$}
   \label{fig50}
 \end{minipage}
 \hfill
 \begin{center}
 \begin{minipage}[b]{0.65\textwidth}
   \includegraphics[width=\textwidth]{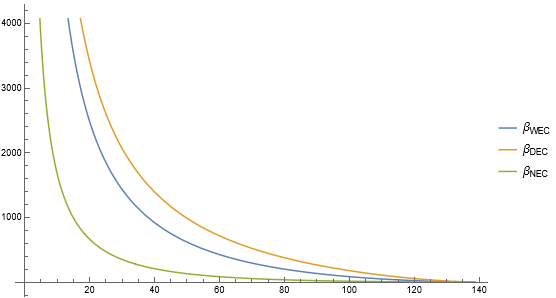}
   \caption{$\beta_{\text{WEC}}, \beta_{\text{DEC}}, \beta_{\text{NEC}}$ vs $\gamma_0$ for vanishing $\rho^m,\, \rho^m-p^m$ and $\rho^m+p^m$}
   \label{fig51}
 \end{minipage}
 \end{center}
\end{figure}

In this case $\gamma_0<138.6$ and $\gamma_0\neq 0$. From figure \ref{fig51}, we observe that NEC, WEC, DEC are satisfied in the range 
\[\beta<\frac{2158.3120\left(138.6-\gamma_0\right)^{\frac{3}{2}}}{2.27\gamma_0^2+164.934\gamma_0}).\] 
SEC is satisfied for 
\[\beta\ge\frac{2917.72\left(138.6-\gamma_0\right)^{3/2}}{18.711\gamma_0-\gamma_0^2}.\] Take note of the change of inequality when the denominator is negative for certain values of $\gamma_0$.
\newpage
\begin{center}
    {\underline{Case 2}: \boldsymbol{$\gamma=\gamma_1 a(t)$}}
\end{center}
\begin{align}
    \kappa \rho^m=&\frac{1.27445\beta\left(2.73\gamma_1^2-239.778\gamma_1\right)}{\left(69.3-\gamma_1\right)^{3/2}}+14407.5\,,\\
    \kappa (\rho^m-p^m)=&26221.6-\frac{8.03752\beta\gamma_1}{\sqrt{69.3-\gamma_1}}\,,\\
    \kappa (\rho^m+p^m)=&\frac{0.849632\beta\left(-1.27\gamma_1^2-63.756\gamma_1\right)}{\left(69.3-\gamma_1\right)^{3/2}}+2593.34\,,\\
    \kappa (\rho^m+3p^m)=&\frac{5.09779\beta\left(88.011\gamma_1-2\gamma_1^2\right)}{\left(69.3-\gamma_1\right)^{\frac{3}{2}}}-21034.9\,.
\end{align}
\begin{figure}[H]
\hfill
 \begin{minipage}[b]{0.49\textwidth}
   \includegraphics[width=\textwidth]{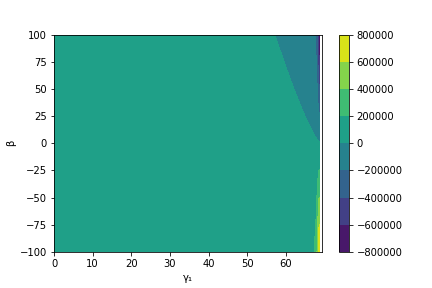}
   \caption{$\kappa \rho^m$}
   \label{fig52}
 \end{minipage}
 \hfill
 \begin{minipage}[b]{0.49\textwidth}
   \includegraphics[width=\textwidth]{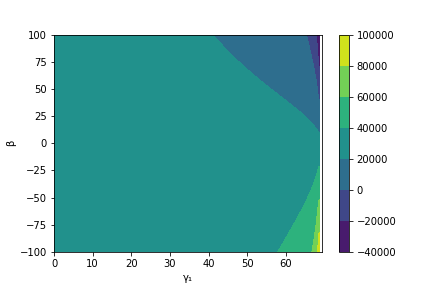}
   \caption{$\kappa (\rho^m-p^m)$}
   \label{fig53}
 \end{minipage}
 \hfill
 \begin{minipage}[b]{0.49\textwidth}
   \includegraphics[width=\textwidth]{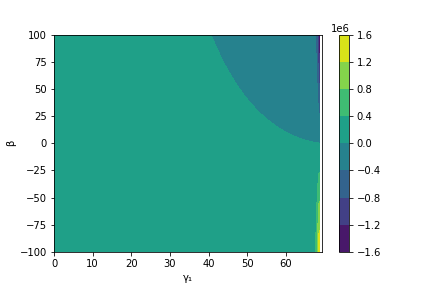}
   \caption{$\kappa (\rho^m+p^m)$}
   \label{fig54}
 \end{minipage}
 \hfill
 \begin{minipage}[b]{0.49\textwidth}
   \includegraphics[width=\textwidth]{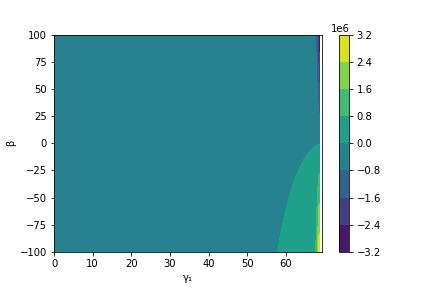}
   \caption{$\kappa (\rho^m+3p^m)$}
   \label{fig55}
 \end{minipage}
 \end{figure}
 \hfill
 \begin{figure}[H]
 \begin{center}
 \begin{minipage}[b]{0.65\textwidth}
   \includegraphics[width=\textwidth]{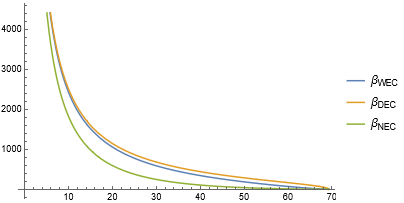}
   \caption{$\beta_{\text{WEC}}, \beta_{\text{DEC}}, \beta_{\text{NEC}}$ vs $\gamma_1$ for vanishing $\rho^m,\, \rho^m-p^m$ and $\rho^m+p^m$}
   \label{fig56}
 \end{minipage}
 \end{center}
\end{figure}

In this case $\gamma_1<69.3$ and $\gamma_1\neq 0$, from figure \ref{fig56}, we observe that NEC, WEC, DEC are satisfied in the range 
\[\beta<\frac{3052.31\left(69.3-\gamma_1\right)^{3/2}}{1.27\gamma_1^2+63.756\gamma_1}).\]
SEC is satisfied for 
\[\beta\ge\frac{4126.28\left(69.3-\gamma_1\right)^{\frac{3}{2}}}{88.011\gamma_1 -2\gamma_1^2}.\] Take note of the change of inequality when the denominator is negative for certain $\gamma_1$.
\begin{center}
{\underline{\bf{Case 3}: \boldsymbol{$\gamma=\gamma_2 H(t)$}}}
\end{center}
\begin{align}
    \kappa \rho^m=&\frac{15.0039\beta\gamma_2\left(1.7158\gamma_2-1.84\right)}{\left(2-0.73\gamma_2\right)^{\frac{3}{2}}}+14407.5\,,\\
    \kappa (\rho^m-p^m)=&\frac{20.0052\beta\gamma_2\left(2.2487\gamma_2-4.3442\right)}{\left(2-0.73\gamma_2\right)^{\frac{3}{2}}}+26221.6\,,\\
    \kappa (\rho^m+p^m)=&-\frac{10.0026\beta\left(-0.65\gamma_2-3.1684\right)\gamma_2}{\left(2-0.73\gamma_2\right)^{\frac{3}{2}}}+2593.34\,,\\
    \kappa (\rho^m+3p^m)=&-\frac{60.0156\beta\left(0.5329\gamma_2-2.5042\right)\gamma_2}{\left(2-0.73\gamma_2\right)^{\frac{3}{2}}}-21034.9\,.
\end{align}
\begin{figure}[H]
 \begin{minipage}[b]{0.49\textwidth}
   \includegraphics[width=\textwidth]{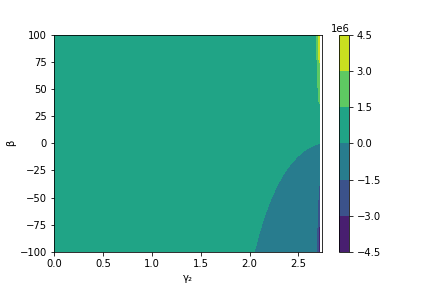}
   \caption{$\kappa \rho^m$}
   \label{fig57}
 \end{minipage}
 \hfill
 \begin{minipage}[b]{0.49\textwidth}
   \includegraphics[width=\textwidth]{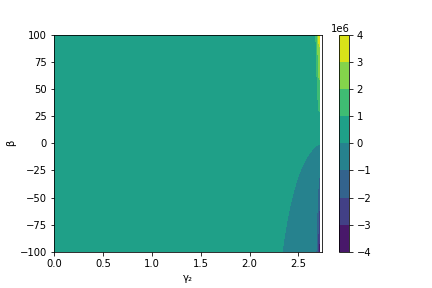}
   \caption{$\kappa (\rho^m-p^m)$}
   \label{fig58}
 \end{minipage}
 \hfill
 \begin{minipage}[b]{0.49\textwidth}
   \includegraphics[width=\textwidth]{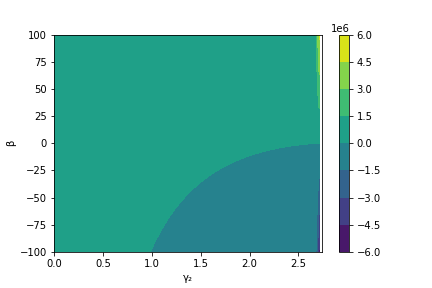}
   \caption{$\kappa (\rho^m+p^m)$}
   \label{fig59}
 \end{minipage}
 \begin{minipage}[b]{0.49\textwidth}
   \includegraphics[width=\textwidth]{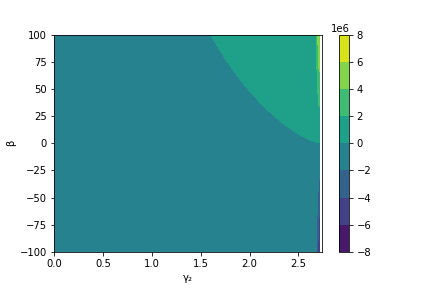}
   \caption{$\kappa (\rho^m+3p^m)$}
   \label{fig60}
 \end{minipage}
  \end{figure}
 \begin{figure}[H]
 \begin{center}
 \begin{minipage}[b]{0.65\textwidth}
   \includegraphics[width=\textwidth]{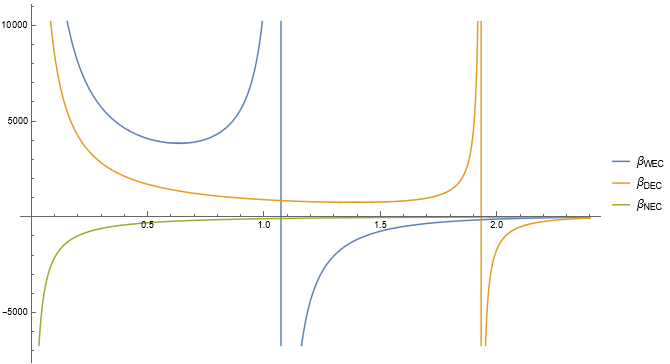}
   \caption{$\beta_{\text{WEC}}, \beta_{\text{DEC}}, \beta_{\text{NEC}}$ vs $\gamma_2$ for vanishing $\rho^m,\, \rho^m-p^m$ and $\rho^m+p^m$}
   \label{fig61}
 \end{minipage}
 \end{center}
\end{figure}

In this case $\gamma_2<2.74$ and $\gamma_2\neq 0$. From figure \ref{fig61}, we observe that NEC, WEC, DEC are satisfied in the range 
\[\frac{259.267\left(2 -0.73\gamma_2\right)^{\frac{3}{2}}}{\left(-0.65\gamma_2-3.1684\right)\gamma_2}<\beta<-\frac{1310.74\left(2-0.73\gamma_2\right)^{\frac{3}{2}}}{\gamma_2\left(2.2487\gamma_2-4.3442\right)}.\] 
SEC is satisfied for 
\[\beta\le-\frac{350.491\left(2-0.73\gamma_2\right)^{\frac{3}{2}}}{\left(0.5329\gamma_2-2.5042\right)\gamma_2}.\] Take note of the change of inequality of the denominator for certain $\gamma_2$.
\begin{center}
    \underline{\bf{Case 4}: \boldsymbol{$\gamma=\frac{\gamma_4}{a}$}}
\end{center}

\begin{align}
   \kappa \rho^m=&0.0826703\gamma _4\beta+14407.5\,,\\
    \kappa (\rho^m-p^m)=&0.110227\gamma _4\beta+26221.6\,,\\
    \kappa (\rho^m+p^m)=&0.0551135\gamma _4\beta+2593.34\,,\\
    \kappa (\rho^m+3p^m)=&-21034.9\,.
\end{align}
\begin{figure}[H]
 \begin{center} 
 \begin{minipage}[b]{0.65\textwidth}
   \includegraphics[width=\textwidth]{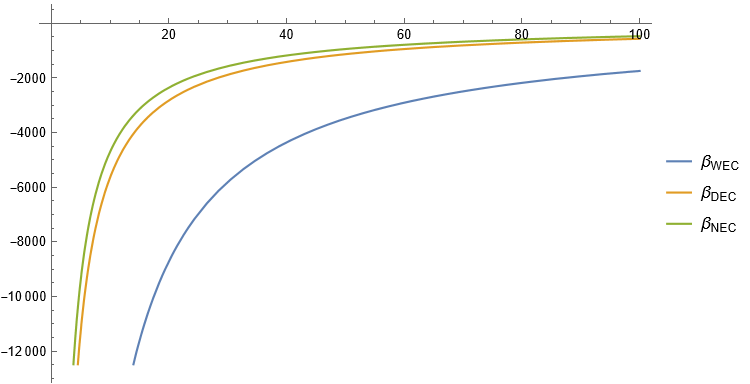}
   \caption{$\beta_{\text{WEC}}, \beta_{\text{DEC}}, \beta_{\text{NEC}}$ vs $\gamma_4$ for vanishing $\rho^m$, $\rho^m-p^m$ and $\rho^m+p^m$}
   \label{fig204}
 \end{minipage}
 \end{center}
\end{figure}
From FIG (\ref{fig204}), we observe that WEC, DEC and NEC are satisfied for $\beta\gamma_4>-47054.5$, in the region above the green line. whereas the SEC remains constant.
\section{Concluding remarks}\label{sec3}
We have investigated the three possible classes of affine connections compatible with the symmetric teleparallel structure in the spatially flat Friedmann-Lema\^itre-Robertson-Walker (FLRW) spacetime. Each connection depends on an unknown and so far unrestricted time-varying parameter $\gamma(t)$. Considering ordinary barotropic fluid as the matter source, we have derived the pressure and energy equations for the ordinary matter and their effective counterparts and noticed that only for the connection type I, the parameter $\gamma(t)$ does not enter into the cosmological dynamics. In our present study three ansatz of $\gamma(t)$ are chosen; a constant function $\gamma(t)=\gamma_0$, $\gamma(t)\propto a(t)$ and $\gamma(t)\propto H(t)$. Moreover, a special choice of  $\gamma(t)\propto a^{-3}(t)$ and $\gamma(t)\propto a^{-1}(t)$ is used in Connection class II and III, respectively, to receive the non-metricity scalar value $Q=-6H^2$, same as in the connection class I (which is independent of $\gamma(t)$).

The $f(Q)$ theory of gravity provides an alternative way to explain the late-time accelerating state of the universe without invoking either the existence of an extra spatial dimension or an exotic component of dark energy. However, each $f(Q)$ model giving rise to a new gravity theory forces us to decide how to filter viable models on physical grounds. The classical ECs are an ideal candidate to constrain the parameters of $f(Q)$ models. Two of the most popular $f(Q)$ models, namely, $f(Q)=Q+\beta Q^2$ and $f(Q)=Q+\beta \sqrt{-Q}$ are considered and the range of ($\beta,\gamma$) for valid ECs are computed for each class of connections. Observational values of some cosmological parameters in the late-time era are used for this purpose, yielding an effective equation of state $\omega^{eff}=-0.82$.

As mentioned earlier, for Connection I, the non-metricity scalar, energy and pressure terms are independent of $\gamma(t)$, and we observe that depending merely on the positivity or negativity of the model parameter $\beta$, all the ECs are satisfied for both models. The other connections are not so trivial, and the parameter $\gamma(t)$ engages in the dynamics. We have plotted all the required linear combinations of pressure and energy density against the parameters ($\beta,\,\gamma$) and the permitted (for valid ECs) ranges for each case have been provided. 

Since the usual coincident gauge based discussions for Connection I of $f(Q)$ theory in the spatially flat FLRW metric in Cartesian coordinates cannot distinguish itself from the well-studied $f(T)$ theory, it does not make sense to repeat already known cosmological characteristics from the same set of Friedmann equations in a novel theory of gravity. Therefore, the future cosmologists must look into these new connections (II and III) investigated here. Our analysis firmly shows a novel path including a technically simple route to tackle the unrestricted parameter $\gamma(t)$ and atleast two viable gravity models with a definite range for the model parameter.


\end{document}